\documentclass[12pt,preprint]{aastex}

\def\MSUN{\rm M$_{\odot}$}

\shorttitle{Chandra Imaging of NGC 1365} \shortauthors{Wang et al.}

\begin{document}

\title{Imaging the Circumnuclear Region of NGC 1365 with Chandra}


\author{Junfeng Wang, G. Fabbiano, M. Elvis, and G. Risaliti\altaffilmark{1}}
\affil{Harvard-Smithsonian Center for Astrophysics, 60 Garden St, Cambridge, MA 02138} \email{juwang@cfa.harvard.edu; pepi@cfa.harvard.edu; elvis@cfa.harvard.edu; risaliti@cfa.harvard.edu}
\altaffiltext{1}{Current Address: INAF-Arcetri Observatory, Largo E, Fermi 5, I-50125 Firenze, Italy}

\author{J. M. Mazzarella, J. H. Howell, and S. Lord}
\affil{Infrared Processing and Analysis Center, California Institute of Technology, MS 100-22, Pasadena, CA 91125}
\email{mazz@ipac.caltech.edu; jhhowell@ipac.caltech.edu; lord@ipac.caltech.edu}

\begin{abstract}

We present the first {\em Chandra}/ACIS imaging study of the
circumnuclear region of the nearby Seyfert galaxy NGC 1365. The X-ray
emission is resolved into point-like sources and complex, extended
emission. The X-ray morphology of the extended emission shows a
biconical soft X-ray emission region extending $\sim$5 kpc in
projection from the nucleus, coincident with the high excitation
outflow cones seen in optical emission lines particularly to the
northwest. Harder X-ray emission is detected from a kpc-diameter
circumnuclear ring, coincident with the star-forming ring prominent in
the {\em Spitzer} mid-infrared images; this X-ray emission is
partially obscured by the central dust lane of NGC 1365. Spectral
fitting of spatially separated components indicates a thermal plasma
origin for the soft extended X-ray emission ($kT=0.57$ keV). Only a
small amount of this emission can be due to photoionization by the
nuclear source. Detailed comparison with [OIII]$\lambda5007$
observations shows the hot interstellar medium (ISM) is spatially
anticorrelated with the [OIII] emitting clouds and has thermal
pressures comparable to those of the [OIII] medium, suggesting that
the hot ISM acts as a confining medium for the cooler photoionized
clouds. The abundance ratios of the hot ISM are fully consistent with
the theoretical values for enrichment from Type II supernovae,
suggesting that the hot ISM is a wind from the starburst circumnuclear
ring.  X-ray emission from a $\sim450$ pc long nuclear radio jet is
also detected to the southeast.

\end{abstract}



\section{Introduction}\label{intro}

Understanding active galactic nucleus (AGN)--galaxy interaction and
feedback is of great consequence for the study of both galaxy and AGN
evolution (e.g., Silk \& Rees 1998; Kauffmann et al. 2003; Hopkins et
al. 2006). The intense ionizing radiation, relativistic jets, and
winds from the AGNs interact with the interstellar medium (ISM) of
their host galaxies, and may either stimulate or suppress star
formation (e.g., Sazonov, Ostriker \& Sunyaev 2004). Visible
signatures of direct AGN-host interaction include kpc-scale regions
with [OIII] and H$\alpha$ line emission, known as the extended narrow
line region (ENLR), which have been found in many nearby Seyfert
galaxies (Schmitt et al. 2003; Veilleux et al. 2003).  Soft X-ray
emission is seen associated with the ENLR, providing an opportunity to
obtain X-ray diagnostic of the physical properties of the interacting
ISM (e.g., Young et al. 2001; Yang et al. 2001; Ogle et al. 2000,
2003; Bianchi et al. 2006; Evans et al. 2006; Kraemer, Schmitt \&
Crenshaw 2008).  Another important link between AGNs and starbursts is
that molecular gas accreted to the nuclear region may induce a
starburst (Elmegreen 1994), feed a central AGN, or both. AGN-related
shocks may further stimulate star formation (e.g., Scoville 1992;
Gonz{\'a}lez Delgado et al.\ 1998; Veilleux et al.\ 1995; Veilleux
2001).  Star formation in the disks and galactic winds in hot halos of
spiral galaxies can be traced by the diffuse X-ray emission as well
(e.g., Tyler et al. 2004, Strickland et al. 2004; Swartz et al. 2006;
Warwick et al. 2007).  NGC 1365, the object studied in this paper, is
ideal for these investigations, since it hosts an AGN and also has IR
emission coincident with active, intensive nuclear star formation.

\begin{figure}[H]
\epsscale{0.7}
\plotone{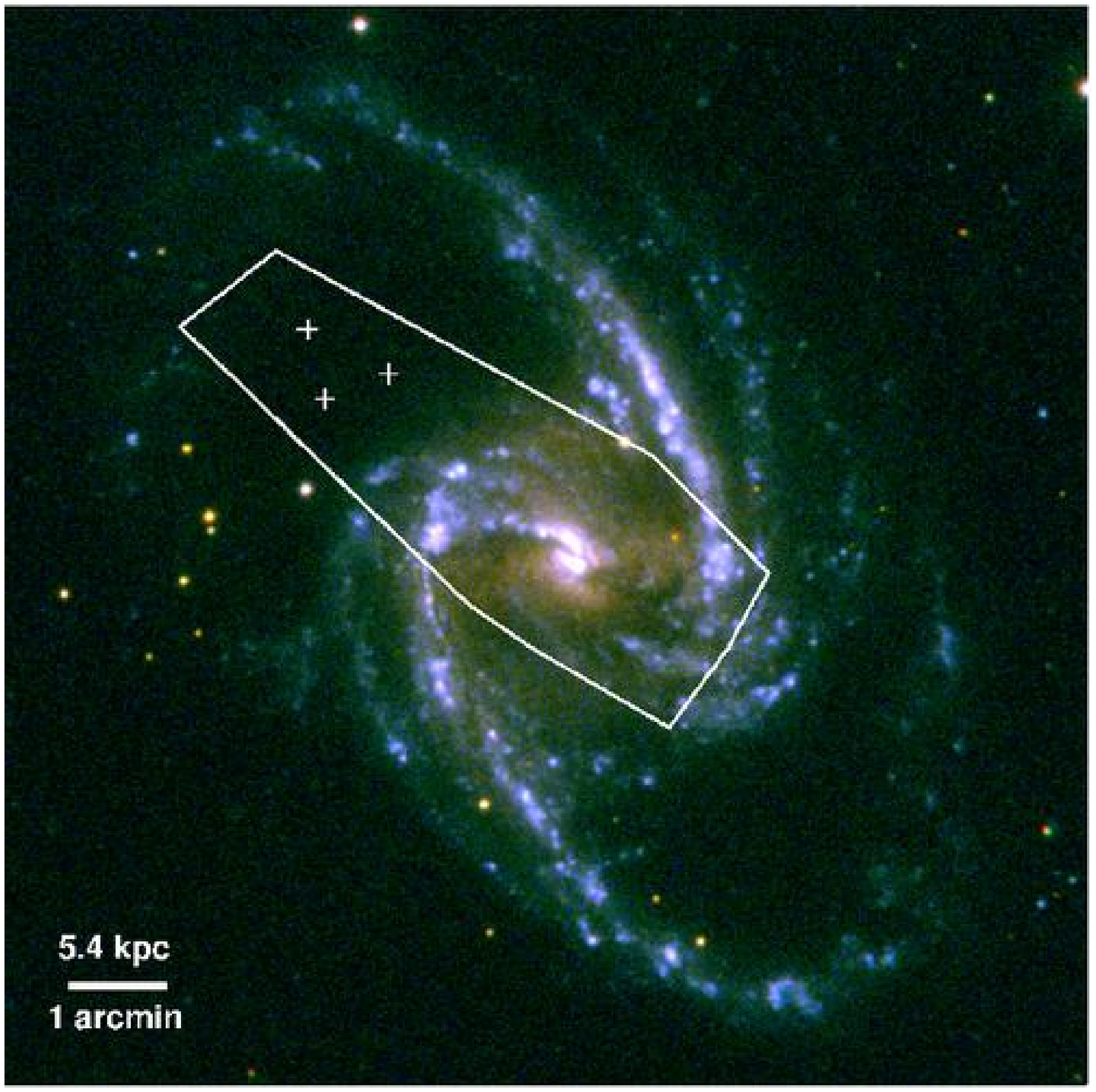}
\caption{Composite ultraviolet ($UV$) image of NGC 1365 from {\em XMM}–Newton/OM. Red corresponds to the $U$ filter, green to the $UVW1$ filter and blue to the $UVM2$ filter. The area covered by all seven Chandra fields is outlined by a white polygon. The centers of background extraction regions are marked with plus symbols (see text). The image is $\sim10\arcmin\times 10\arcmin$. North is up and east to the left.
\label{om}}
\end{figure}

NGC 1365 is a nearby ($D=18.6\pm 1.9$ Mpc, $1\arcsec \sim 90$ pc;
Madore et al. 1998; Silbermann et al. 1999) archetype barred spiral
galaxy (SBb(s)I; Sandage \& Tammann 1981), and likely a member of the
Fornax cluster (Jones \& Jones 1980). A thorough review of the NGC
1365 galaxy is given in Lindblad (1999). It hosts vigorous star
formation in the circumnuclear region and a variably obscured Seyfert
1.5 nucleus\footnote{Other classifications exist in the
  literature. For example, Turner, Urry \& Mushotzky (1993) and
  Maiolino \& Rieke (1995) classified the NGC 1365 nucleus as a
  Seyfert 2 and 1.8, respectively.} (Veron et al. 1980; Hjelm \&
Lindblad 1996; Risaliti et al. 2007). The spectacular symmetric main
spiral arms extending from the ends of a strong bar are apparent in the
{\em XMM}-Newton Optical Monitor (OM) images (Figure~\ref{om}). Dark
dust lanes run across the nuclear region and partially obscure the
nucleus.

NGC 1365 has long been known to exhibit a biconical outflow (Burbidge
\& Burbidge 1960; Phillips et al. 1983; Storchi-Bergmann \& Bonatto
1991). A $\sim 5\arcsec$ long ($\sim$450 pc) 100$^{\circ}$ wide
conical [OIII]$\lambda5007$ emission line region (ELR) is present to
the southeast (SE) and to the northwest (NW) of the nucleus (Hjelm \&
Lindblad 1996; Kristen et al. 1997; Veilleux et al. 2003), which
suggests an AGN ionization cone, although a starburst driven outflow
explanation has also been advanced (Komossa \& Schulz 1998). In the
Hubble Space Telescope ({\em HST}) images, the inner $3\arcsec$ of the
conical outflow seen in [OIII] is resolved into a number of small
clouds and larger agglomerations (Kristen et al. 1997).  A high
excitation [NII]$\lambda6583$/H$\alpha$ region is aligned with the
[OIII] structure, but appears to be rectangular (Veilleux et
al. 2003). Sandqvist et al. (1995) show that the inner parts of the
[OIII] region contain a nuclear radio jet, extending $5\arcsec$ SE
along the galaxy minor axis.  Note that the spatial properties of the
NGC 1365 galaxy and the outflow cones are well determined from these
multiwavelength studies (see Lindblad 1999 review and references
therein): the position angle (PA) of the line of nodes is
$\approx220^{\circ}$, and the inclination angle of the galaxy is
$40^{\circ}$; the [OIII] cone extends from the nucleus out of the
galactic plane with its symmetry axis closely aligned with the
rotation axis of the galaxy, with a full opening angle of about
100$^{\circ}$; the counter cone to the NW is partially obscured by the
absorbing dust in the galaxy plane.

Another prominent characteristic of NGC 1365 is the circumnuclear
star-forming ring with a diameter of $\sim 14\arcsec$ ($\sim 1.3$ kpc;
Kristen et al. 1997). The kpc-size nuclear vicinity contains bright
optical ``hot spots'' (Sersic \& Pastoriza 1965; clearly visible in
the UV images shown in Figure~\ref{om}). The circumnuclear regions are
resolved into many compact super star clusters (Kristen et al. 1997),
forming an elongated ring surrounding the AGN. Its minor axis is
parallel to both the bi-cone axis and the galaxy minor axis,
suggesting the possibility of a collimating torus (Sandqvist 1999).
High resolution mid-infrared imaging of the ring (Galliano et
al. 2005) also unveils a circumnuclear population of point-like
sources coincident with bright CO spots (Sakamoto et al. 2007) and
non-thermal radio continuum sources (Sandqvist et al. 1995; Forbes \&
Norris 1998), which are interpreted as embedded young massive star
clusters. The biconical ELR/outflow and the circumnuclear ring should
be readily detectable in the {\em Chandra} images as hinted from the
spatially extended X-ray emission in the ROSAT HRI image (Stevens et
al. 1999).

In this paper we present a high spatial resolution X-ray imaging study
of the complex circumnuclear region of the active galaxy NGC 1365 with
the {\it Chandra} X-ray Observatory. NGC 1365 has received a great
deal of attention as a result of X-ray observing campaigns to monitor
the obscuration of the AGN (e.g., Risaliti et al. 2005a,b; Risaliti et
al. 2007). Comparing to observations of NGC 1365 from previous
generation X-ray satellites and the {\em XMM}-Newton observatory, the
{\em Chandra} images offer an unprecedented view of the innermost
regions of NGC 1365, which is critical to disentangle the different
emission components.

\section{Observations and Data Reduction}\label{obs}

NGC 1365 was first imaged in 2002 with the back-illuminated chip of
the Advanced CCD Imaging Spectrometer spectroscopy array (ACIS-S;
Garmire et al. 2003) positioned at the focal point of {\it Chandra
  X-ray Observatory} High Resolution Mirror Assembly (van Speybroeck
et al. 1997; Weisskopf et al. 2002). It was further monitored with six
{\em Chandra}/ACIS-S observations in ``1/4 window'' mode during April
2006, resulting in a total exposure time of $\sim$100 ks (see
Table~\ref{obslog} for details; the overlap region common to these
observations is outlined in Figure~\ref{om}).

The data products were analyzed with the {\it Chandra} X-Ray Center
(CXC) CIAO v4.0 software and HEASOFT v6.4 package\footnote{See
  \url{http://cxc.harvard.edu/ciao/} and
  \url{http://heasarc.gsfc.nasa.gov/lheasoft/} for more
  information.}. The new release of calibration files
CALDB\footnote{\url{http://cxc.harvard.edu/caldb/}} v3.4.3 were used.
The data were processed by the CXC (ASCDS version 7.6.7-7.6.9);
verification and validation of the data products showed no
anomalies. Following the standard {\em Chandra} ACIS data preparation
thread, we reprocessed all level 1 data to create new level 2 events
for consistency.  A preliminary run of CIAO tool {\em wavdetect}
(Freeman et al. 2002) is done for point source detection in the
central $8\arcmin$ region of each observation. A $10\arcmin$ radius
circle centered at the S3 aimpoint was used to query the Naval
Observatory Merged Astrometric Dataset (NOMAD) catalogue (Zacharias et
al. 2004) and the extracted optical sources are searched for
counterparts to {\em Chandra} detections. Using optical point sources
in the field detected in X-rays, the absolute astrometry for
Observation ID (ObsID) 3554 is measured to be accurate to within
$0\farcs5$. By comparing the positions of the bright point like X-ray
sources in common between observations, we then corrected the absolute
astrometry for any relative offsets between the 2006 data sets and
ObsID 3554. The relative shifts were $\sim1$ pixel.

To remove periods of high background, for each ObsID we extracted a
light curve for chip S3, excluding any bright sources in the
field. The light curves for 2006 observations showed no strong
flares. However, a period of enhanced background was present in the
last quarter of ObsID 3554, and 1.6 ks exposure of affected data was
screened out.  Finally all seven observations were merged into a
single event file after correcting the relative astrometric
offsets. Detailed study of the point source population is deferred to
a future paper (J. Wang et al. in preparation), which will also use
the {\em XMM}-Newton observations of NGC 1365.

In addition to {\em Chandra} data, ancillary UV and mid-IR images were
also used in this paper.  NGC 1365 was observed with the {\em Spitzer
  Space Telescope} (Werner et al. 2004) as part of a survey of
luminous infrared galaxies (LIRGs; $L_{IR} \ge
10^{11}L_{\odot}$)\footnote{\url{http://goals.ipac.caltech.edu/spitzer/Spitzer.html}}. Observations
with the Infrared Array Camera (IRAC; Fazio et al. 2004) were dithered
to correct for cosmic ray events and bad pixels, and raster mapping
was used to construct mosaics of the galaxy at 3.6, 4.5, 5.8 and 8.0
$\mu$m (Spitzer AOR \#12346880).  Details of the IRAC observations and
data processing are described by Mazzarella et al. (2008, in
preparation). The XMM-Newton OM data were aquired from {\em
  XMM}-Newton ObsID \#0205590301 (PI: G. Fabbiano), using the
Processing Pipeline System (PPS) products processed with Science
Analysis System (SAS; version 6.5.0).  The UV images shown in Figure~1
used the following OM filters (Mason et al. 2001): $U$ ($\lambda_0\sim
3472\AA$), $UVW1$ ($\lambda_0\sim 2905\AA$), and $UVM2$
($\lambda_0\sim 2298\AA$).

\section{The X-ray Color Images and Morphology}\label{truecol}

Smoothed images of the NGC 1365 region were first created without
removing the point sources. This enabled us to examine the spatial
distribution of the point sources with respect to the galaxy. We
extracted images from the merged data in the commonly used soft band
(0.3--0.65 keV), the medium band (0.65--1.5 keV), and the hard-band
(1.5--7 keV). Note that the medium band encompasses the emission from
the Fe-L blend. The 0.3 keV low-energy boundary is the lowest energy
where ACIS remains well calibrated, and the high energy 7 keV cut-off
is chosen to limit the contribution from the background.

\begin{figure}[H]
\epsscale{0.7}
\plotone{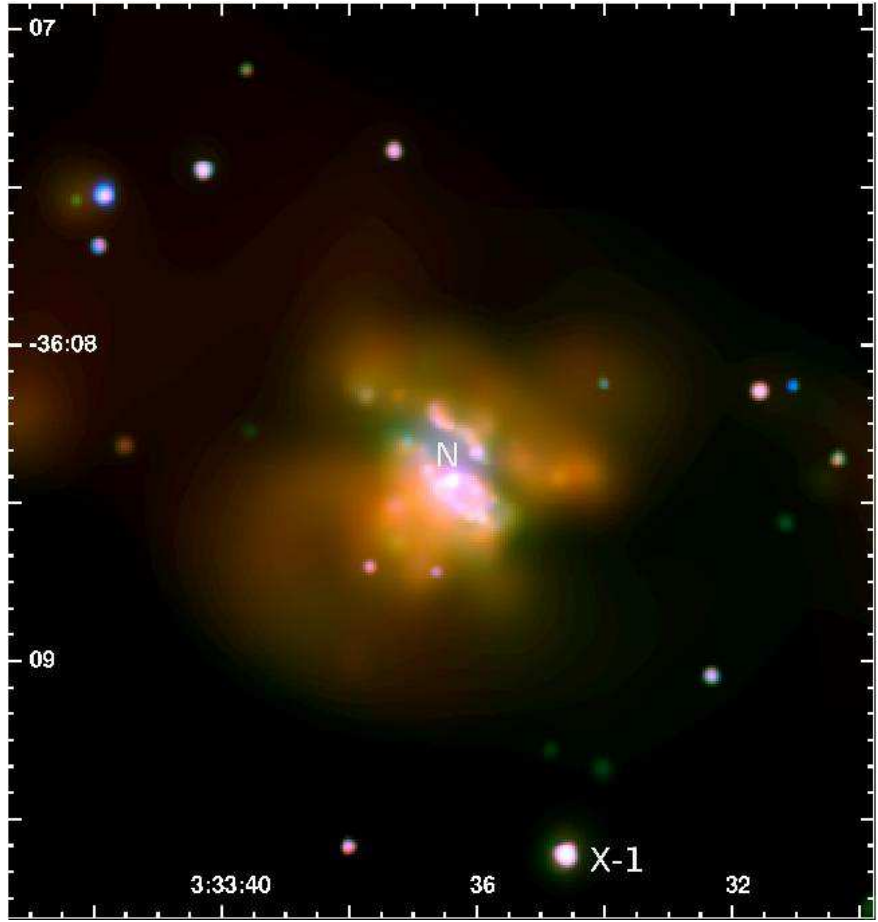}
\caption{Adaptively smoothed image showing point sources and diffuse emission in the inner $3\arcmin\times3\arcmin$ region of NGC 1365.  Red represents soft-band X-ray emission (0.3--0.65 keV), green represents medium-band X-ray emission (0.65--1.5 keV), and blue for hard-band emission (1.5--7.0 keV).
\label{3c_diff}}
\end{figure}

To derive exposure-corrected images, exposure maps were created for
individual observations and bands, then reprojected to create the
combined exposure maps matching the merged data. The band-limited
images were then divided by the appropriate exposure maps and
adaptively smoothed. This process enhances faint extended features
while preserving the high resolution of brighter features. In order to
prevent large variations in the adaptive smoothing scales between the
bands with different counts, we applied to the soft and hard bands the
same smoothing scales calculated for the medium band image, which has
the best signal to noise ratio (S/N). The significance was chosen
between 2.4 and 5$\sigma$ above the local background (corresponding to
the smoothing Gaussian kernel of varying scales between 1 and 40
pixels).  Figure~\ref{3c_diff} presents the ``false color'' composite
image of the central $3\arcmin\times 3\arcmin$ region of NGC 1365,
where the soft, medium, and hard band smoothed images are shown in
red, green, and blue, respectively.  Besides the bright nucleus
(marked as ``N'' in Figure~\ref{3c_diff}), some 40 point sources are
present.  The bright off-nuclear X-ray source towards the lower right
corner is a well-studied, variable ultraluminous X-ray source (ULX)
NGC 1365 X-1 (Komossa \& Schulz 1998), most likely a
50--150\MSUN\ accreting black hole (Soria et al. 2007).

\begin{figure}[H]
\epsscale{0.5}
\plotone{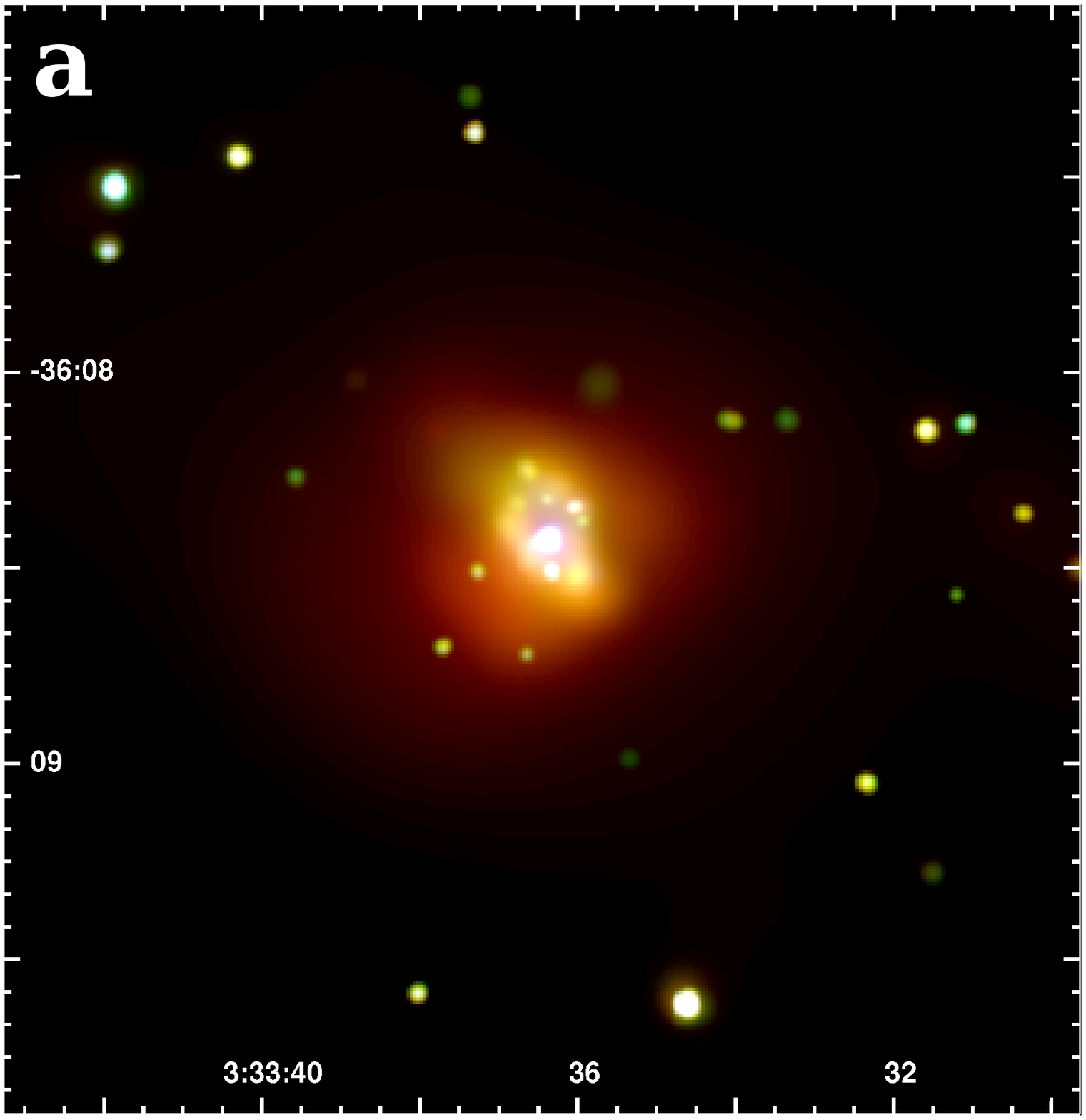}
\plotone{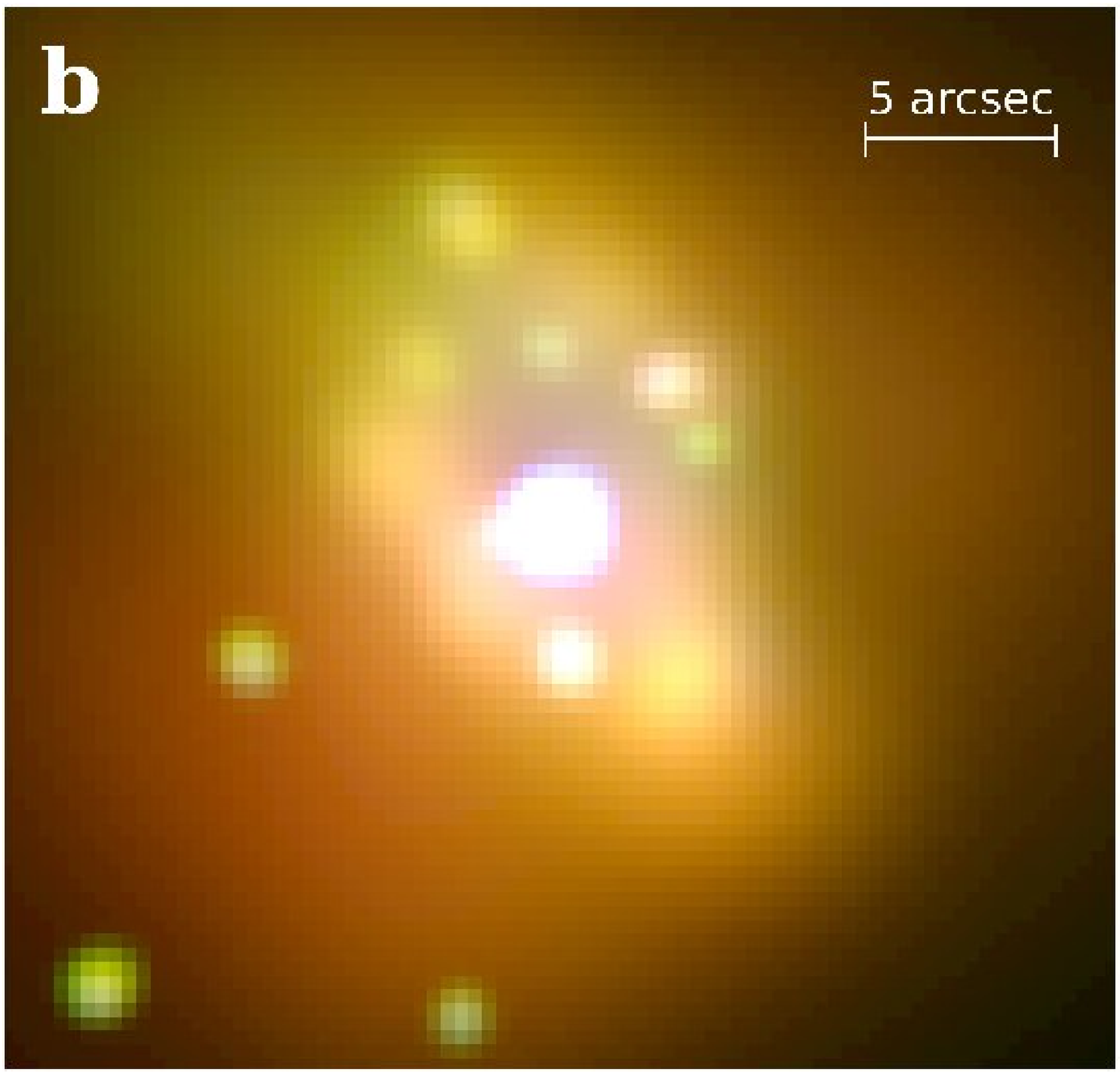}
\caption{(a) Adaptively smoothed image of the same region in
  Figure~\ref{3c_diff}, but with a different energy--color
  mapping. Here red refers to 0.3--1.5 keV emission, green is 1.5--3.0
  keV, and blue 3.0--7.0 keV. (b) Same as (a) but zoomed-in to the
  inner $30\arcsec\times 30\arcsec$ region showing the circumnuclear
  ring.
\label{3c_ring}}
\end{figure}

Figure~\ref{3c_diff} shows the circumnuclear ring (mainly pink; $\sim
14\arcsec$ in diameter), intersected by the obscuring dust lane
(blue). This ring is emphasized with an alternative set of energy
bands shifted to higher energies images in Figure~\ref{3c_ring}, in
which the soft band (red) is 0.3--1.5 keV, the medium band (green) is
1.5--3.0 keV, and the hard band (blue) is 3.0--7.0 keV; the same
smoothing was applied as for Figure~\ref{3c_diff}.  The hard emission
(presumably from young supernovae remnants and X-ray binaries) is less
affected by the obscuration from the dust lane, tracing the extent of
the star forming ring. Note that with this energy band selection the
(broader) soft band now has the best $S/N$ of the three bands and so the
smoothing scales from this band were used to smooth the images in the
other two bands. This alternate band selection is only used here in
this paper.

\begin{figure}[H]
\epsscale{0.7}
\plotone{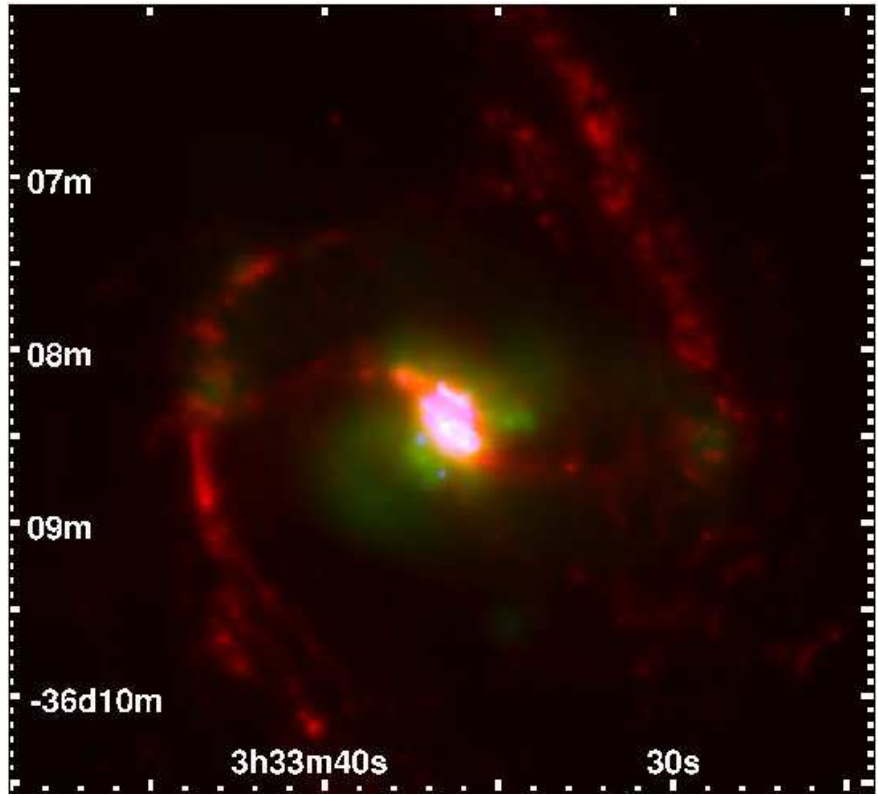}
\caption{A $4\arcmin\times4\arcmin$ {\em Spitzer} mid-IR and {\em Chandra} X-ray composite image of NGC 1365. Red is 8$\mu$m mid-IR emission, green is medium-band X-ray emission, and blue is hard-band X-ray emission.
\label{spitzer}}
\end{figure}

\begin{figure}[H]
\epsscale{0.6}
\plotone{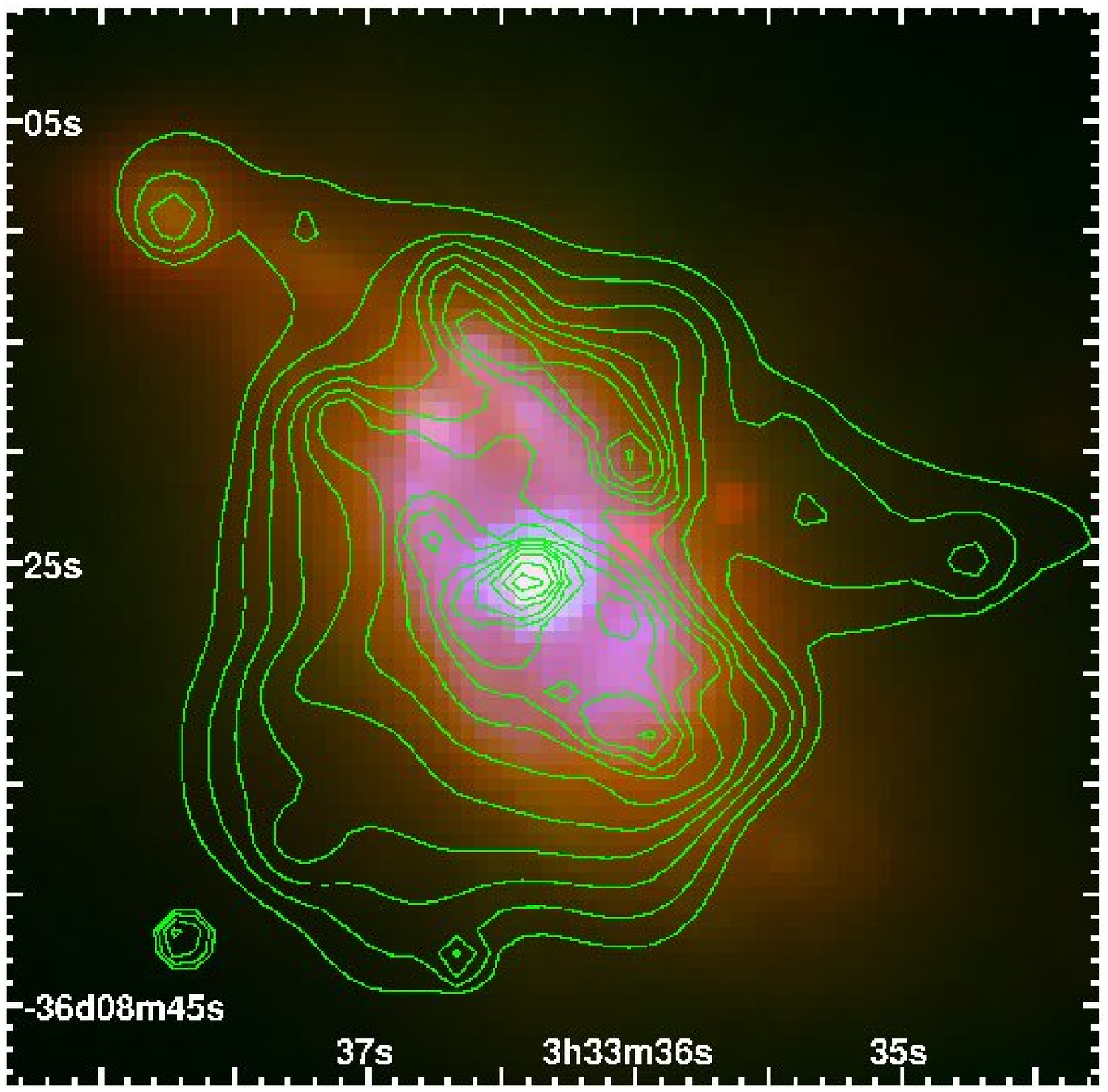}
\plotone{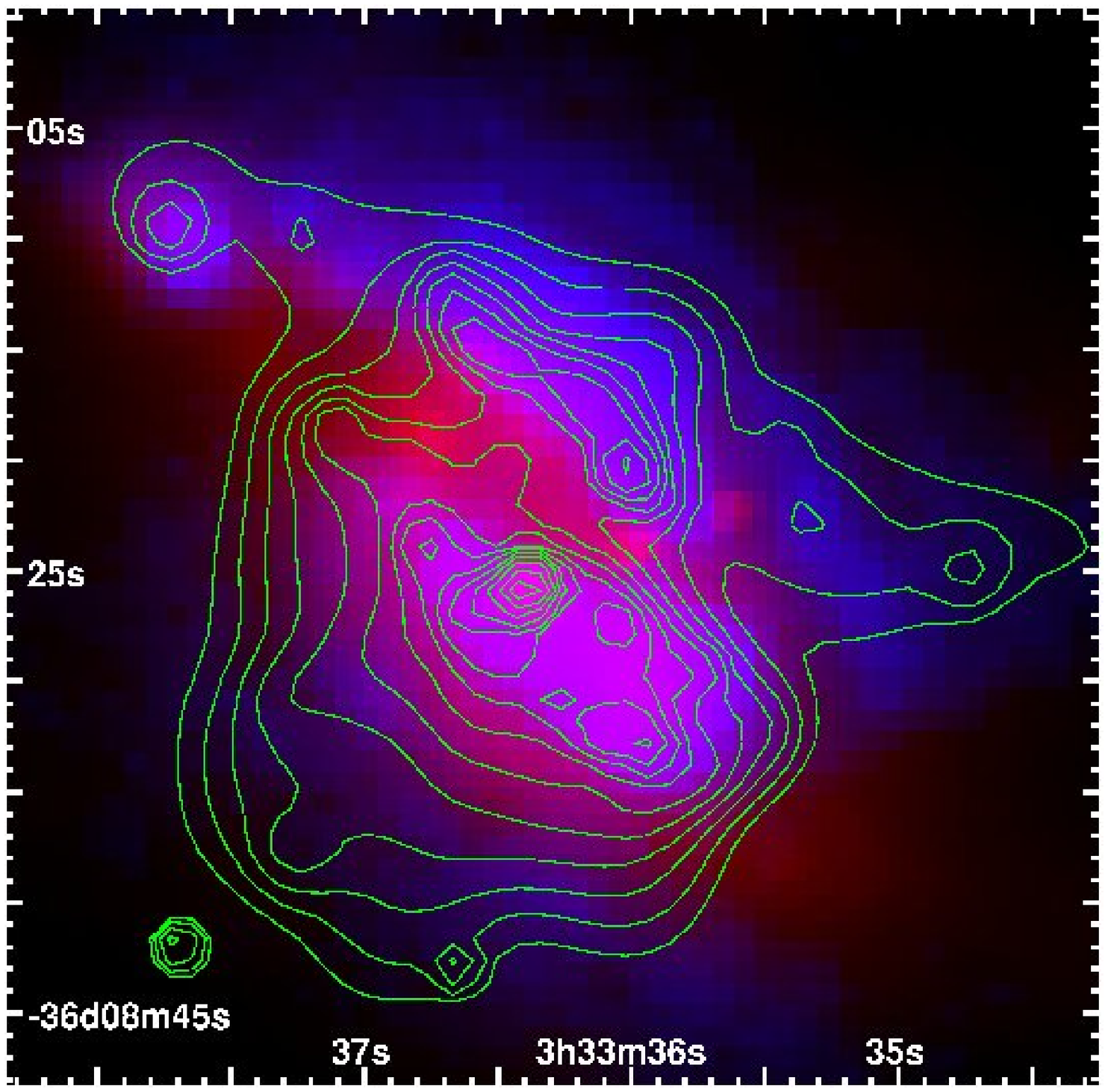}
\caption{Top: {\em Spitzer}/IRAC composite image of the central 1 arcmin region of NGC 1365.  Red is 8$\mu$m emission, green is $4.5\mu$m, and blue $3.6\mu$m.   Bottom: Composite mid-IR/UV image of the same region.  Red is {\em Spitzer}/IRAC 8$\mu$m image, and blue is $XMM$-Newton/OM $U$-band image.  The X-ray contours (medium-band, without removing point sources) are shown in both panels to facilitate comparison. 
\label{spitzer2}}
\end{figure}

\begin{figure}[H]
\epsscale{0.7}
\plotone{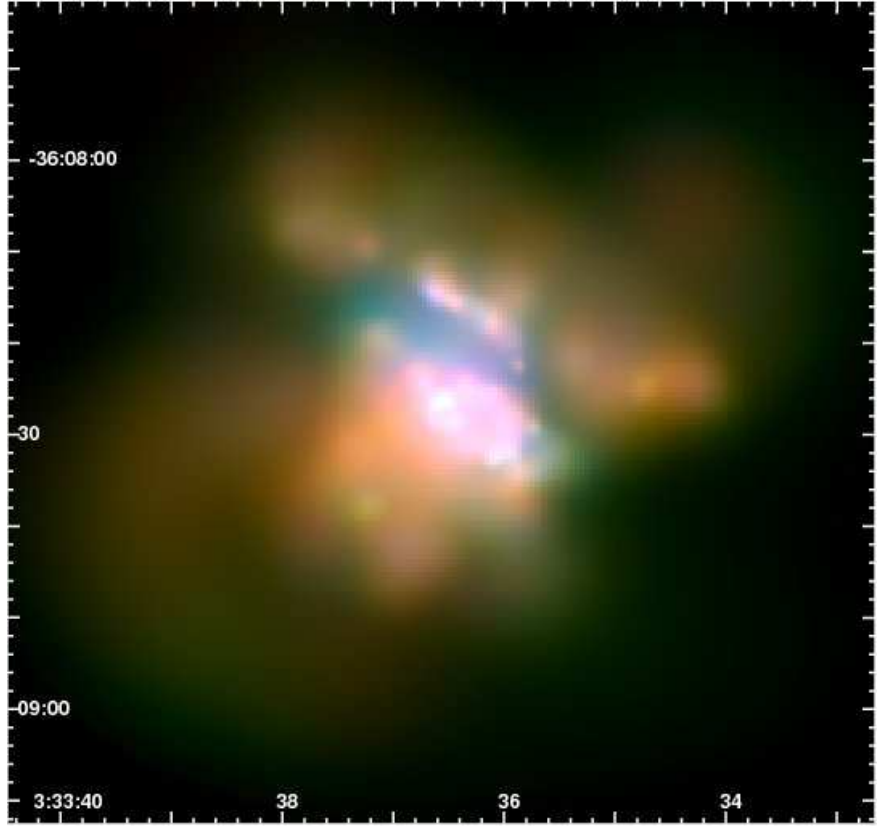}
\caption{Adaptively smoothed image of the diffuse emission in NGC 1365. Point source have been removed. Two conical structures (with $\sim 100^{\circ}$ opening angles) are seen extending towards NW (top right) and SE (lower-left) from the center.  Red represents soft-band X-ray emission (0.3--0.65 keV), green medium-band X-ray emission (0.65--1.5 keV), and blue hard-band emission (1.5--7.0 keV).
\label{truecolor}}
\end{figure}

Comparison with {\em Spitzer} data clearly relates the hard X-ray ring
to the circumnuclear star-forming ring prominent in the infrared.
Figure~\ref{spitzer} shows the {\em Chandra} X-ray image of the
diffuse emission (from Figure~\ref{truecolor}) in NGC 1365
(green/blue), together with the mid-IR (8~$\mu$m) view of the galaxy
(red). Figure~\ref{spitzer2} zooms in to the central 1 arcmin, which
shows a composite mid-IR image of the circumnuclear ring and a
composite mid-IR/UV image of the same region with an overlay of the
medium-band X-ray emission.

The circumnuclear ring is compact and maintains regularity close to
the center. Such circumnuclear star formation rings are commonly
observed in strongly barred spirals and assumed to be dynamically
associated with the Inner Lindblad Resonance (ILR)---inflowing gas
accumulates to form a massive star forming ring between the two inner
resonances (see Lindblad 1999).  The Inner ILR and Outer ILR in NGC
1365 is $\sim3\arcsec$ and $\sim30\arcsec$ from the nucleus,
respectively (Lindblad et al. 1996), hence the ring is located $\sim
4\arcsec$ outside of the Inner ILR.  We note a hot-spot located $\sim
22\arcsec$ ($\sim 2$ kpc) NE of the nucleus, which shows co-spatial
X-ray, UV, and mid-IR emission.  IR filaments bridge the ring and this
object, suggesting that it is related to the galaxy, instead of being
a background object, and may be associated with the bar.  The spatial
correlation between the X-ray and UV emission is apparent. Because of
the obscuring dust lane, which is traced by its bright IR emission, a
bright ridge of X-ray/UV emission with hot-spots appears coincide with
the edge of the IR ring $\sim 10\arcsec$ (1 kpc) N and NW of the
nucleus, probably related to the discrete regions of active star
formation associated with the ring.

To study the large scale diffuse emission, we followed the procedure
of Baldi et al. (2006a) to create mapped-color images with point
sources removed. Given the short individual exposure times, we
performed point source detection in the combined exposures in the full
band (0.3--7.0 keV). The point-like ACIS detections including the
nucleus were then removed from the data, and the resulting holes were
filled by interpolating the surrounding background emission. The same
procedure to define source regions and background regions employed by
Baldi et al.(2006a) was followed. The results were visually compared
with the band-limited images and the smoothed images to ensure that no
apparent individual point sources were missed. In the circumnuclear
ring region, a few ``point'' sources detected by {\it wavdetect}
appear extended. These could be either association of unresolved point
sources or bright patches and knots of diffuse emission. To check
whether they affect the spectral analysis of the circumnuclear ring,
we created separate event files without removing these sources with
uncertain classification. These event files were used in spectral
extraction and fitting described later in this paper.

The images with point sources removed and holes filled in the three
bands (0.3--0.65 keV; 0.65--1.5 keV; and 1.5--7 keV) were adaptively
smoothed, then combined to create a ``mapped-color'' image of the
diffuse emission shown in Figure~\ref{truecolor}. The mapped color
image shows complex morphology and striking substructures that have
not been seen in lower resolution X-ray images previously.

The overall morphology of the X-ray emission is rather consistent with
the optical picture described in \S~\ref{intro}:

\begin{enumerate}

\item Extended soft X-ray emission appears to resemble a bicone (NW
  and SE) structure with the apex at the nucleus. The SE structure is
  more extended ($\sim0.8\arcmin$ measured from the nucleus) and
  visually closer to a cone-shape. Its inner region is brighter and
  appears to contain substructure. The NW cone is only bright at the
  base close to the nuclear region ($\sim0.4\arcmin$), with fainter
  emission extends towards far NW. Both cones are incomplete, possibly
  due to obscuring material in the galaxy plane and bulge.  The
  measured projected half opening angle of the cone in the X-ray image
  is approximately $\sim55^{\circ}$. The X-ray cones appear to be
  edge-brightened, which is consistent with the hollow cones seen in
  the optical (Lindblad 1999).

\item Harder X-ray emission is associated with the circumnuclear
  star-forming ring, which appears pink and clumpy in the
  image.

\item The hardest X-ray emission, which appears as a $\sim
  4\arcsec$-wide blue band intersecting the top portion of the
  circumnuclear ring, is coincident with the dust extinction lane of
  the galaxy.

\end{enumerate}

\section{The Line-Strength Map}

A quick examination of the spectra extracted from the diffuse emission
show the presence of the Oxygen+Iron+Neon (O+Fe+Ne) emission blend
(0.6--1.16~keV), the Magnesium (Mg)-XI (1.27--1.38~keV) and the
Silicon (Si)-XIII (1.75--1.95~keV) lines (Figure~\ref{spectra}). To
identify regions of strong apparent line emission and to select
possible anomalous abundance regions for spectral analysis, we
generated an emission line map, following Fabbiano et al.\ (2004) and
Baldi et al. (2006a), to highlight regions of prominent emission lines
in the hot ISM.

Three band-limited images were created to encompass the lines noted
above. A ``continuum template'' was extracted using the 1.4--1.65~keV
and 2.05--3.05~keV bands which are not visibly contaminated by
emission lines, and was used to determine the underlying continuum in
the emission line bands.  All point-like sources were removed and the
holes were filled with background as described in
\S~\ref{truecol}. The O+Fe+Ne 0.6--1.16~keV band image contains the
largest number of counts and so was used to set the smoothing scales
for the other line and continuum images.

To determine the continuum contribution in each line image, we
extracted the spectrum of the entire diffuse emission of NGC 1365 and
fitted the continuum with a simple two-component thermal
bremsstrahlung model, excluding the three energy ranges with strong
line emission. The best fit model provides the appropriate scaling
factors between the continuum level in our continuum template and the
continuum in a given line band. The scaling factors are 2.5, 0.2, and
0.19 for the 0.6--1.16~keV, 1.27--1.38~keV, and 1.75--1.95~keV band,
respectively. Each line image was continuum subtracted after weighting
the continuum image by these scaling factors, and the resulting images
were combined to create a mapped-color line-strength image
(Figure~\ref{line}).

Figure~\ref{line} shows that the emission-line structure of the hot
ISM in NGC 1365 is also quite complex. Silicon emission (blue) appears
most pervasive in the dust lane; this could be due to obscuration of
the softer energy photons in this region, and this possibility is
explored further below. Magnesium lines (green) and O+Fe+Ne emission
(red) are prevalent in most regions. Part of the NW diffuse region
shows enhanced O+Fe+Ne emission.  Noticeably there is an area (orange
in color) south of the star-forming ring that is lacking in silicon
line emission.

\section{Spectral Analysis}\label{specanal}

\subsection{Data Extraction}\label{specfit}

\begin{figure}[H]
\epsscale{0.7}
\plotone{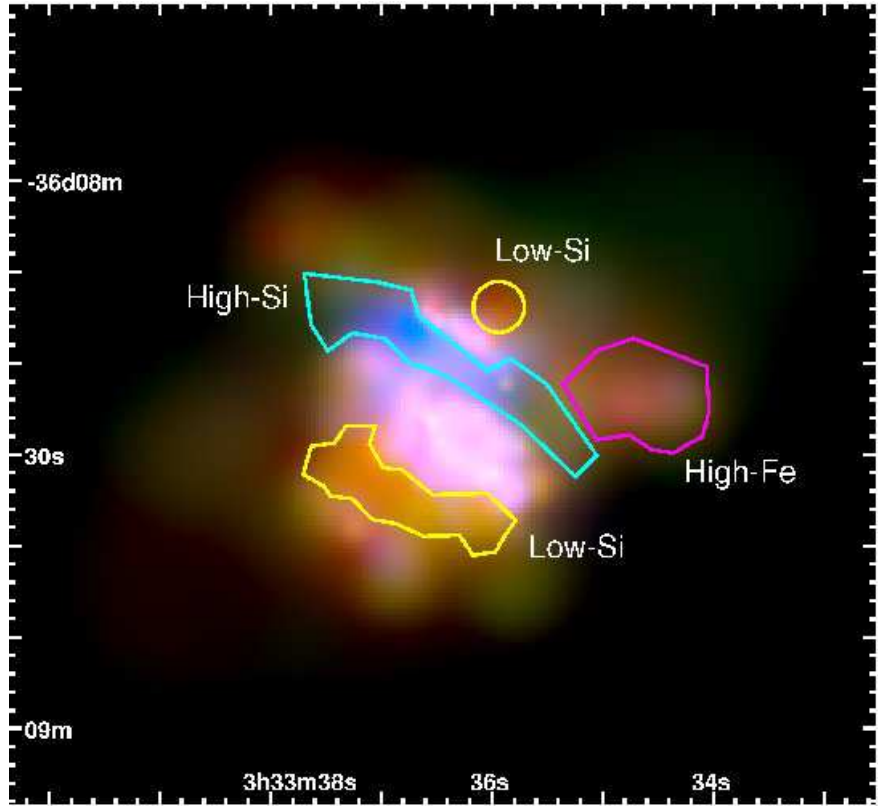}
\caption{Adaptively smoothed image of the X-ray emission-line images in NGC
  1365.  Red represents Oxygen+Iron+Neon (O+Fe+Ne) emission blend
  (0.6--1.16~keV), green represents Magnesium (Mg)-XI (1.27--1.38~keV)
  line, and blue represents Silicon (Si)-XIII (1.75--1.95~keV). The
  overlaid polygons outline the extreme line ratio regions for further
  spectral analysis.
\label{line}}
\end{figure}

The mapped-color X-ray image (Figure~\ref{truecolor}) and the
line-strength map (Figure~\ref{line}) demonstrate that the diffuse
emission of the hot ISM in NGC 1365 is rich in spatial and spectral
features. These figures suggest structure in the hot ISM, possibly the
effect of varying absorption columns, plasma temperatures, and metal
enrichment.

\begin{figure}[H]
\epsscale{0.7}
\plotone{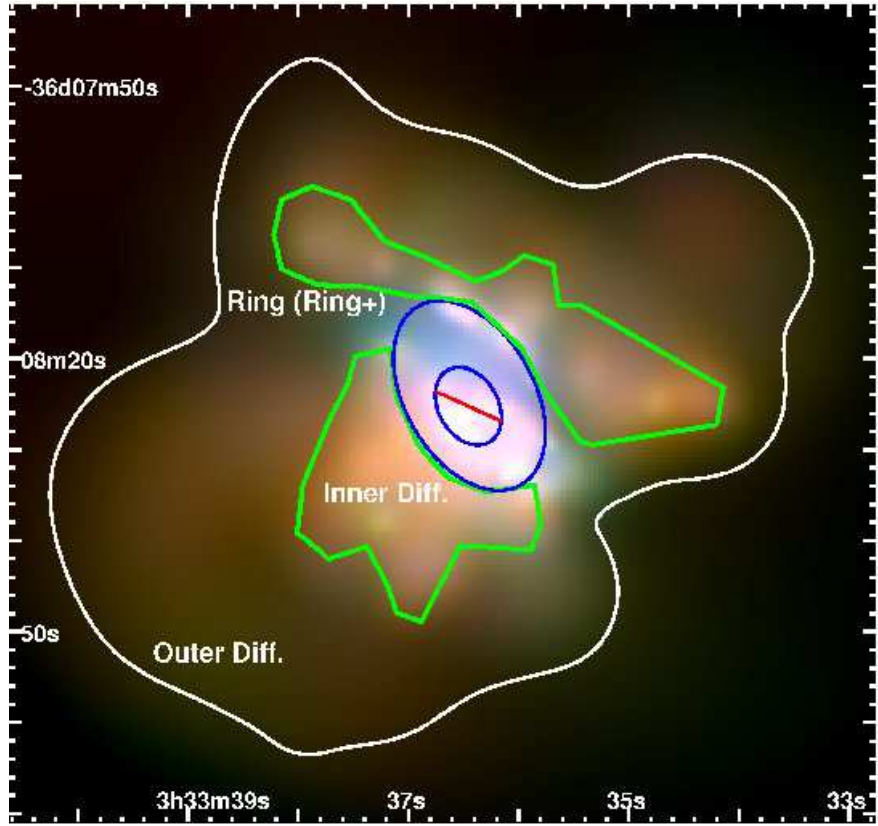}
\caption{Diffuse emission image overlaid with the regions used for
  spectral extraction.  The ``ring'' is outlined by the blue ellipse
  excluding the central nucleus.  The region ``diffuse'' refers to all
  the diffuse emission, excluding the area inside the outer limit of
  the ring.  It is further divided to ``inner diffuse'' (the bright
  portion of the cones; green polygons) and ``outer diffuse'' (the remaining faint diffuse
  emission).
\label{c_diff_reg}}
\end{figure}

As outlined in Figures~\ref{c_diff_reg} and \ref{line}, we identified
five separate regions based on morphology and three regions based on
apparent abundance pattern, to perform spectral analysis of the X-ray
emission from the hot ISM. The region ``ring'' refers to the
circumnuclear ring, excluding the nucleus (see \S~\ref{truecol} and
Figure~\ref{3c_ring}). The region ``diffuse'' refers to all the
diffuse emission, excluding the area inside the outer limit of the
ring.  It is further divided into ``inner diffuse'' (the bright portion
of the cones) and ``outer diffuse'' (the faint diffuse emission). In
addition, for the ring area we also extracted photons from the
``ring'' region including the {\it wavdetect} selected point sources
that appear extended (see \S~\ref{truecol}). We dubbed the region and
the corresponding extraction ``ring+''.  We did not further divide the
diffuse emission into sub-regions (e.g., the ring is separated by the
dust lane into north-ring and south-ring) to have good statistics for
the spectral fitting.  Regions related to the emission line features
noted in Figures~\ref{line} were also created, with descriptive names
(``high-Si'', ``low-Si'', and ``high-Fe'').  We emphasize here that,
even though the extended emission is referred to as ``diffuse'', the
X-ray emission is not $bona fide$ diffuse plasma. There is evidence
for extensive clumpiness and X-ray hot-spots. The ``diffuse'' X-rays
include a mixture of hot ISM and unresolved point sources (XRB/SNR)
associated with regions of very active star formation.  The hard power
law component seen in the spectral fit (see \S~\ref{fitting}) is
likely related to the XRB/SNR components.

We extracted spectra separately from the individual ObsIDs, for each
diffuse emission regions described above. We excluded all the point
source areas above $3\sigma$ described in \S~\ref{truecol}.  We also
extracted background counts representative of the field from three
circular source-free regions (each with a $0.4\arcmin$ radius,
resulting in a total area of 1.5 arcmin$^2$) well outside the galaxy,
but within the ACIS-S3 chip in all seven observations
(Figure~\ref{om}). A new CIAO script $specextract$\footnote{See CIAO
  thread \url{http://cxc.harvard.edu/ciao/threads/specextract/}.}  was
used to extract the region spectra for multiple observations. This
script extracts the spectrum and creates area-weighted Response Matrix
Files (RMF) and Ancillary Response Files (ARF) for each region. For
each region of interest, all source spectra and background spectra
were co-added and the response files were combined with their
appropriate weights using the FTOOLS
software\footnote{\url{http://heasarc.gsfc.nasa.gov/docs/software/ftools/}}. The
resulting spectra were then grouped so as to have signal to noise
ratio $S/N\ge 10$ in each energy bin.

\subsection{Thermal and Power Law Models}\label{fitting}

The extracted spectra were analyzed with XSPEC v12.0 (Arnaud 1996),
using combinations of absorbed optically-thin thermal emission and
power law components to fit the data. We used the Astrophysical Plasma
Emission Code (APEC) thermal-emission model (Smith et al.\ 2001) that
utilizes improved atomic data from the Astrophysical Plasma Emission
Database\footnote{\url{http://cxc.harvard.edu/atomdb/}}.  The
absorption column was modeled with $tbabs$ (Wilms et al.\ 2000), which
includes updated photoionization cross section and abundances of the
ISM, as well as a treatment of interstellar grains and the H$_2$
molecule.

We first considered absorbed single-temperature APEC models as well as
two-temperature APEC models. A single line-of-sight absorption column
was used in all cases, equivalent to Galactic line-of-sight column
($1.4\times 10^{20}$ cm$^{-2}$; Kalberla et al. 2005) combined with
the intrinsic absorption within NGC 1365, which was free to vary.
When the fit required unmeasurably small absorption ($N_H\ll 10^{20}$
cm$^{-2}$), we froze $N_H$ at the Galactic column. As the spectrum of
the integrated diffuse emission clearly shows prominent MgXI
($\sim$1.3 keV), OVII-NeIX/X-Fe-L complex (0.6-1.2 keV), and SiXIII
($\sim$1.8 keV) line features, the abundances of these elements were
left to vary freely. When fitting with the two component models, the
Oxygen abundance was fixed at the solar value as it is poorly
constrained in our data and consistent with solar abundance. A power
law component is introduced to the single-temperature model to allow
for the possible integrated emission of unresolved X-ray binaries that
are too faint to be detected individually. A simple absorbed power law
model give poor fits ($\chi^2_{\nu}>4$) to all the extracted spectra.
To fit the spectra of the three extreme line ratio regions
(Figures~\ref{line}), we only allow the abundances of Mg, Si, and Fe
to vary, given the limited number of extracted counts.

For each of the regions used in the spectral analysis, the net counts
(background subtracted) in the full band, the reduced $\chi^2$
($\chi^2$ over the numbers of degrees of freedom $dof$,
$\chi^2_{\nu}$), the resulting best-fit temperature(s) $kT$ (keV),
absorbing column density $N_H$ (cm$^{-2}$), and power law photon index
$\Gamma$ (if a power law component is invoked) are summarized in
Table~\ref{mytable1}, together with the abundances of Oxygen, Neon,
Magnesium, Silicon, and Iron relative to the solar values.  The last
column of Table~\ref{mytable1} gives the adopted ``best-fit model'',
which is chosen based on having the smallest $\chi_{\nu}^2$. If more
than one model can fit the spectrum well and the $\chi_{\nu}^2$ values
are close (difference less than 0.1), a simpler model with less free
parameters is preferred. The errors are quoted at $1.65\sigma$ (90\%
confidence interval) for one interesting parameter.  The spectra of
the regions, together with the best-fit models and the
data$-$(best-fit-model) residuals, are shown in Figure~\ref{spectra}.

In summary, these results show variations in different morphological/physical regions: 

\begin{enumerate}
\item The large scale diffuse emission (``diffuse'') is well fitted with a
single temperature thermal model, with a well determined
$kT=0.57^{+0.05}_{-0.03}$ keV. The absorption may be a little in
excess of the Galactic column towards NGC 1365 ($\Delta
N_H=2.2^{+4.7}_{-2.2}\times 10^{20}$ cm$^{-2}$). To adequately fit the
emission lines in the ACIS spectrum, the data require sub-solar
abundances ($Z/Z_{\odot}\sim 0.2-0.4$).  The sub-regions of the
diffuse emission show similar characteristics to the combined
emission.  The absorption column of the outer diffuse emission
(``outer diff'') is low, consistent with just Galactic absorption,
which is reasonable, as the region is further away from the spiral
disk.

\item The spectra of the circumnuclear ring (and ``ring+'') require
higher absorption columns, $\log N_H\sim 21.2$ cm$^{-2}$, and a hard
spectral component in addition to the $kT=0.6\pm 0.05$ keV thermal
component similar to the diffuse emission (``diffuse'').  The large
$N_H$ is not surprising given the dust lane and higher obscuration
towards the nuclear region. In the power-law fit this hard component
has a photon index $\Gamma\sim 2.5$; using a two-temperature thermal
model, we obtain a hard $kT\sim 3$ keV, but the fit with power-law
component has a better $\chi^2$ for the same numbers of free
parameters.  In either case a hard component is required, and maybe
due to the unresolved X-ray source population in the ring. The
best-fit model suggests super-solar abundances in the hot ISM ($\sim3$
times $Z_{\odot}$).

\item For the extreme line ratio regions, all the fits have
  $\chi^2_{\nu}$ less than 1 indicating the models are over complex
  for the data. Nevertheless, the ``high-Si'' region (dust lane) shows
  high absorption column ($\log N_H\sim 21.5$) as expected. The fit
  suggests under-abundant Fe, but this result may be affected by the
  obscuration of the dust lane.  Taking the $N_H=3.2\times 10^{21}$
  cm$^{-2}$ (Table~\ref{mytable2}) and a visual extinction $A_V\sim 3$
  mag (Roy \& Walsh 1997; Kristen et al. 1997; Galactic contribution
  is removed from $N_H$ and $A_V$), we measured a gas-to-dust ratio
  $N_H/A_V \sim 10^{21}$ cm$^{-2}$ mag$^{-1}$, which is roughly in
  agreement with the standard Galactic gas-to-dust ratio
  $N_H/A_V=1.9\times 10^{21}$ cm$^{-2}$ mag$^{-1}$ (Maiolino et
  al. 2001 and references therein). The extracted spectra of
  ``low-Si'' region also suggests extremely low abundance of Si
  ($Z_{Si}/Z_{\odot}\sim 10^{-3}$ preferred from the fit; we consider
  it unconstrained). We note that a similar low Si abundances have
  been reported in some regions of The Antennae (NGC 4038/9; Baldi et
  al 2006b).  A possible explanation for this deficiency is that a
  large fraction of the silicon has cooled and locked in dust
  grains. SN Ia enrichment may also yield a relatively low Si
  abundance (Nagataki \& Sato 1998).

\item Lastly, for all the spectral fitting, we emphasize that the
temperature and absorption column for the same region determined from
various models show little variation, indicating that these parameters
are rather robust.

\end{enumerate}

In Table~\ref{lumin} we list the derived emission parameters from the
spectral analysis of each region. From these emission parameters and
those listed in Table~\ref{mytable1}, following Fabbiano et al. (2003)
and Baldi et al. (2006b), we can further estimate some physical
properties of the emitting plasma, such as electron density $n_e$
($\approx n_H$), thermal energy content $E_{th}$, cooling time
$\tau_c$, pressure $p$, hot ISM mass $M_{ISM}$, and supernovae rate
$R_{SN}$. These quantities are given in Table~\ref{mytable2}. The
emitting volume ($V$) of the hot gas in the region of interest
(assuming a cylindrical geometry) is derived using the projected area
on the plane of the sky multiplied by a height of $200$ pc (for the
typical depth of a spiral disk).  Considering the approximate nature
of the geometry and depth of the hot gas, the inclination of the
galaxy ($\sim40\deg$) has little effect on the estimated emitting
volume. The filling factor $\eta$ is assumed to be 100\% in
Table~\ref{mytable2}, and the derived parameters have weak dependence
on $\eta$ ($n_e, p\propto \eta^{-1/2}$ and $E_{th}, \tau_c,
R_{SN}\propto \eta^{1/2}$).  These estimates assumed an $\eta$ of
unity and the geometry of the X-ray emitting gas, and should not be
treated as precise measurements (see discussion in Baldi et
al. 2006b).

\subsection{Photoionization Model}\label{photo}

Although a thermal model accounts well for the bulk of the diffuse
X-ray emission, some positive residuals are seen around 0.53 keV
(NVII-OVII) and around 0.7--0.8 keV (possibly OVII and OVIII radiative
recombination continua) in the thermal fit of diffuse emission
(Figure~\ref{spectra}), which hint at the presence of additional
possible photoionized, emission lines.  Emission from a photoionized
medium might be the dominant component in the inner region around the
nucleus (Guainazzi et al., in preparation), though not at radii
greater than $10\arcsec$.  To explore the line emission from
photoionized gas, we used
XSTAR\footnote{\url{http://heasarc.gsfc.nasa.gov/docs/software/xstar/xstar.html}}
(Kallman \& Bautista 2001) v2.1ln7 and the XSPEC12 local model
$photemis$\footnote{Available at
  \url{ftp://legacy.gsfc.nasa.gov/software/plasma\_codes/xstar/warmabs21ln7.tar.gz}}
to model the photoionized contribution to the extended X-ray emission
(``diffuse'' and ``inner diffuse''). Using a photoionized model alone
yielded poor fits ($\chi^2_{\nu}\gg 5$) of the spectra (see further
discussed from an energetic viability perspective in
\S~\ref{discuss}).  Unlike in detailed photoionization modeling
studies of high resolution grating spectra (e.g., Ogle et al. 2000,
2003), we froze the parameters in our best-fit thermal model, and
added an emission component from a warm photoionized emitter.  We
assume the gas is photoionized by an incident spectrum with a power
law index $\Gamma=1.7$ (consistent with the average spectral index of
the nucleus, Risaliti et al. 2007), and calculated the photoionized
emission with XSTAR.  We find that the best-fit ionization parameter
is $log\xi=0.5\pm 0.4$ where $\xi=L/(nR^2)$, $L$ is the luminosity of
the ionizing source, $n$ is the ion density, and $R$ is the distance
to the source. We estimate an $E.M.=2\times 10^{61}$ cm$^{-3}$ for the
photoionized emitter, about 1\% of the $E.M.$ of the thermal
component.  Since the X-ray absorbing gas towards NGC 1365 nucleus
shows significant variability, we also attempted modeling with an
incident spectrum of $\Gamma=2.3$ (the spectral slope derived from
high S/N $XMM$-Newton spectra with the lowest absorption column, which
represents the best measurements of the continuum parameters; Risaliti
et al. 2008, in preparation).  The resulting ionization parameter is
poorly constrained, $log\xi=0.2\pm 1.2$, and the photoionized emitter
may contribute $\sim$17\% of the total $E.M.$.

\section{Discussion}

\subsection{The nuclear cone: hot or photoionized ISM?}\label{discuss}

A biconical ENLR is a common feature of nearby Seyfert galaxies
(Schmitt et al. 2003), and the associated soft X-ray emission has been
suggested to originate from the same photoionzed gas seen in [OIII]
(e.g., Ogle et al. 2000; Bianchi et al. 2006).  For the ionization
cones in NGC 1365 (in the ``diffuse'' region), the soft emission is
consistent with a thermal origin, as the thermal model provides
adequate fit to the spectra while the pure photoionization model
fails.  Further morphological evidence from X-ray and [OIII] emission
will be discussed in the next section.  The presence of the Fe-L
complex, which is often seen dominant in spectra from hot gas in
collisional ionization equilibrium (Phillips 1982) supports a thermal
origin, although Ogle et al. (2003) cautioned that these lines could
arise from recombination and photoexcitation as well. {\em Chandra}
observations of NGC 1068 (Ogle et al. 2003) demonstrated that AGN
outflows can show a wide range in ionization ($\log \xi=$1--3). The
derived $\log \xi=0.5\pm 0.4$ (or $\log \xi=0.2\pm 1.2$ if
$\Gamma=2.3$) in our simple photoionization fitting (\S~\ref{photo})
is low but marginally consistent with the lower boundary of $\xi$.

Following Weaver et al. (1995) and Evans et al. (2006), we further
examine the energetic viability of the photoionization model with the
measured X-ray luminosity, assuming that the extended soft X-ray
emitting gas is predominantly photoionized by the nuclear emission.
In this case, an ionizing parameter $\xi=100$ ergs cm s$^{-1}$ is
required to produce the strong Fe-L emission feature arising from
ionized gas (Kallman \& McCray et al. 1982; Kallman 1991). Adopting an
emissivity $j(\xi)=10^{-24}$ ergs cm$^3$ s$^{-1}$ for the photoionized
gas and the observed soft X-ray emission $L_X\sim 10^{40}$ ergs
s$^{-1}$, we derived a density $n\sim 0.2$ cm$^{-3}$. It is worth
noting that this density is in excellent agreement with the $n_e\sim
0.1-0.2$ cm$^{-3}$ that we derived in the hot gas from the thermal
model (Table~\ref{mytable2}), indicating that our estimate is
reasonable despite the simple assumption made here. Substituting
$\xi$, $n$, and a distance $R=10\arcsec$ (900 pc) in $\xi=L/(nR^2)$,
the luminosity of the nucleus required to photoionize the gas to the
observed level is $L\sim 2\times 10^{44}$ ergs s$^{-1}$. This is
$\sim$100--1000 times higher than the unabsorbed $L_X\sim
10^{41}-10^{42}$ ergs s$^{-1}$ (2--10 keV) reported in Risaliti et
al. (2005b), suggesting that photoionization alone cannot account for
the extended soft X-ray emission unless the ionizing radiation from
the nuclear source is highly anisotropic.  As NGC 1365 is a relatively
low luminosity AGN, that thermal emission is dominant over an
AGN-continuum photoionized gas may not be surprising.  We note that
NGC 1365 is characterized with a surface density of the star formation
rate $\Sigma_{SFR}=3M_{\odot}$ yr$^{-1}$ kpc$^{-2}$ and a mean gas
surface density $\Sigma_{H_2}=1\times 10^3 M_{\odot}$ pc$^{-2}$
(Sakamoto et al. 2007; a $^{12}$CO-to-H$_2$ conversion factor is
adopted as in Kennicutt 1998). These surface densities follow the
Kennicutt-Schmidt law for a sample of starbursts and normal disk
galaxies (Kennicutt 1998).  Wang et al.(2007) derive a domain in the
$\Sigma_{SFR}$--$\Sigma_{H_2}$ plot where the circumnuclear medium is
significantly affected by the AGN radiation, and with a sample of 57
Seyfert galaxies they show evidence that the circumnuclear star
formation of many Seyferts are suppressed by the AGN feedback.  We
find NGC 1365 lies near but outside of the boundary of the AGN
feedback domain, resembling the starbursts well.

Our spectral results are overall consistent with previous X-ray studies of NGC
1365 that found that the extended soft nuclear emission can be well
fitted by a thermal plasma model with a very low local absorbing
column (Fabbiano et al. 1992; Iyomoto et al. 1997; Komossa \& Schulz
1998; Stevens et al. 1999).  We caution that the low resolution CCD
imaging spectroscopy cannot distinguish whether the ``continuum'' is
actually made of unresolved emission lines from photoionized
gas. Higher resolution grating spectra, especially constraints from
the line intensities can shed more light on the AGN-photoionized
emission component (e.g., Guanazzi et al. 2007, Evans et al. 2007,
Longinotti et al. 2007, Kraemer et al. 2008).

Additional evidence in favor of a thermal hot ISM can be gathered by
comparing our results with the [OIII] properties of the circumnuclear region.
Figure~\ref{oiii} compares the spatial correspondance between the
X-ray emission and continuum subtracted [OIII] image from Veilleux et
al. (2003), enabled by the superior arcsec resolution X-ray image and
the excellent astrometry. Both X-ray and [OIII] are present in the
inner $10\arcsec$ region around the nucleus. Most imortantly, the NW
cone shows some complementary structure between the two types of emission
(a ``swiss-cheese''-like morphology): the X-ray emission is weak where
the [OIII] emission is strong.  The hot X-ray emitting gas
appears delimited by the cooler condensation traced by the [OIII]
emission.  This spatial anti-correlation cannot be attributed to
obscuration or projection, and is consistent with a predominantly
thermal origin for the X-ray emission (Elvis et al. 1983, Ogle et
al. 2000), which was also suggested by our spectral analysis
(\S~\ref{specanal}).  

Our data do not have the resolution to perform a detail comparison of
the emission properties of the numerous smaller clouds within
$2\arcsec$ of the nucleus that were resolved in the {\em HST} [OIII]
image (Kristen et al. 1997). However, assuming a temperature of $\sim
10^4$ K and a density of $\sim 100-1000$ cm$^{-3}$ for the [OIII]
emitting clouds, which are typical values observed in the optical
emission line gas (e.g., Osterbrock \& Ferland 2006, Wilson et
al. 1985, Ferruit et al. 1999), we estimate the gas pressure of the
[OIII] clouds to be $\sim 10^{-10}-10^{-9}$ dyne cm$^{-2}$. These
pressures are comparable to (or smaller than) the thermal pressure of
the hot ISM we calculated in Table~5, implying a possible pressure
equilibrium between the X-ray emitting hot gas and the optical
line-emitting cool gas. The X-ray-emitting gas may serve as the hot
phase confining intercloud medium to the NLR clouds (see Elvis et
al. 1983).

\subsection{Hot Gas Properties and Supernovae Enrichment}\label{SN}

As summarized in \S~\ref{fitting}, the results of our spectral fitting
of the diffuse (ENLR cone, ``diffuse'') and of the circumnuclear ring
emission (``ring''), suggest a pervasive hot ISM with temperatures of
about 0.6~keV. Metal abundances are overall larger (super-solar) in
the ring, and become sub-solar at larger radii in the cone region.
This temperature and abundance pattern is similar to that of the hot
ISM in many regions of NGC 4038/9, ``the Antennae'' galaxies (see
Baldi et al. 2006a and b); not surprisingly, the calculated emission
parameters of the hot gas in NGC 1365, such as electron densities,
pressure and cooling times, are also in agreement.  In the Antennae,
the metal abundances of the hot ISM are consistent with metal
enrichment from SN II ejecta.  As the ring in NGC 1365 is undergoing a
starburst, we may similarly expect to see enhanced abundances.

Nakataki \& Sato (1998) show that if the elemental enrichment is from
type II SN ejecta, the expected [Ne/Fe] and [Mg/Fe] values approach
0.3 and [Si/Fe] approaches 0.5 on average. In contrast, the expected
values from type Ia SN is significantly lower.  Thus we can explore
the metal enrichment in NGC 1365 based on this difference.  We
compiled the element abundances relative to Iron in different regions
in Table~\ref{mytable3}. Together we also show the abundance values
from the averaged hot halo of starburst galaxies (Strickland et
al. 2004), the warm Galactic halo (Savage \& Sembach 1996), and Type
Ia, Type II SNe from a variety of theoretical models (Nakataki \& Sato
1998).  Although there are some uncertainties, the morphologically
selected regions in NGC 1365 show [Ne/Fe], [Mg/Fe], and [Si/Fe] ratios
fully consistent with a type II SN-dominated enrichment scenario (see
Figure~\ref{NeMg}).  As the circumnuclear ring is where active star
formation is occurring and where candidate radio SNe are identified
(Galliano et al.2005), it is plausible that recent episodes of star
formation led to the enrichment of the ISM almost exclusively through
SNe~II injection. Notably the ``ring'' (and ``ring+'') region has
abundance ratios matching very well with the theoretically predicted
values. Two SNe have been seen in NGC 1365 (not related to Galliano et
al. radio SNe) in four decades (Lindblad 1999), giving a rate about
six times higher than our estimated $R_{SN}$ in Table~\ref{mytable2}
but with a large Poisson error bar.  The $HST$ optical study by
Kristen et al.\ (1997) estimate a supernovae rate of $10^{-3}$
yr$^{-1}$ for the luminous star clusters in NGC 1365, which is in good
agreement with our derived $R_{SN}$.  

However, we note that these measurements in NGC 1365, similar to X-ray
estimates, only place a lower limit on the actual, intrinsic supernova
rate as they are severely affected by dust extinction. It is well
established that the extinction in LIRGs is high enough to obscure SNe
in their central regions, even at near-IR wavelengths (Van Buren \&
Greenhouse 1994; see also the near-IR SN survey by Mannucci et
al. 2003). FIR studies of starburst galaxies appear to yield a higher
SNe explosion rate assuming that $L_{FIR}$ is powered mainly by star
formation. Using observations of NGC 253 and M 82, Van Buren \&
Greenhouse (1994) derived the relation $R_{SN} = 2.3\times 10^{-12}
L_{FIR}/L_{\odot}$ yr$^-1$.  A similar factor ($2.4\pm 0.1\times
10^{-12}$) was derived by Mannucci et al. (2003) and independently
($2.7\times 10^{-12}$) by Mattila \& Meikle (2001).  Applying this to
NGC 1365's total $L_{FIR}$ predicts a total SNe rate of $\sim$0.2
yr$^{-1}$, which is $\sim$20 times higher than the largest diffuse
region in Table~\ref{mytable2} with $R_{SN} \sim 0.01$ yr$^{-1}$.
Within the ring alone, using our estimate of $\sim 35$\% of the total
$L_{FIR}$ (see \S~\ref{comp2}) implies $R_{SN}\sim 0.07$ yr$^-1$,
which is $\sim$200 times higher than the X-ray estimate in
Table~\ref{mytable2}.  This discrepancy can be attributed to a number
of uncertainties.  First, although we used absorption corrected X-ray
luminosity to account for the attenuation of intrinsic X-ray
luminosity, this absorption correction derived from spectral fitting
is still not precise, especially when the absorption column is high
enough to obscure the soft X-rays.  In comparison, FIR and radio are
known to be reliable quantitative measures of emitted luminosities as
most galaxies become transparent at these wavelengths.  Second, we
estimated $R_{SN}$ assuming that the mechanical energy released in SN
explosions heat the ISM thermal emission, the efficiency of which
depends on the detailed properties of the interstellar environment. A
low efficiency can increase the required $R_{SN}$ by a factor of
100. In addition, the assumed X-ray emitting volume and volume filling
factor are uncertain and difficult to quantify.  Altogether, the X-ray
estimated SN rate should be deemed as a lower limit, and does not
contradict the FIR-derived SN rate.

Some caution is warranted as the above analysis relies upon the
abundances of the emitting plasma being accurately modeled. Measuring
hot gas elemental abundances through X-ray spectral fitting was
recognized to be quite uncertain (especially with the low resolution
spectra of CCDs) because of ambiguities in the fits (e.g., a
degeneracy between the temperature and metallicity), and strongly
dependent on the model choice (see Dahlem, Weaver, \& Heckman 1998;
Dahlem et al. 2000; Strickland et al. 2002, 2004, Baldi et
al. 2006a). These uncertainties are strongly affected by poor spatial
resolution, and spatially resolved spectroscopy is needed to determine
reliable constraints.  However, the agreement of all the regions on a
set of abundances close to the SNe~II values seems unlikely to occur
by chance and suggest that {\em Chandra} has spatially resolved the
major variations.

\subsection{X-ray emission from the 'jet'}\label{comp}

Figure~\ref{jet} shows that the bright nuclear emission in the full
band (0.3--7 keV) shows a clear asymmetry, extending $\sim 4\arcsec$
towards the SE.  The orientation of this jet-like feature emanating
from the nucleus is well aligned with the cone-axis seen at larger
scale. We consider this to be a candidate X-ray counterpart of the
radio jet (NGC 1365:F, radio spectral index -0.94, Sandqvist et
al. 1995).  However a fit to the extracted spectra ($\sim$300 net
counts) suggests an absorbed power law emission with $\Gamma\sim3.9$,
which is unusally soft for a jet (Harris \& Krawczynski 2006).  If the
X-rays and radio emission are produced in synchrotron radiation from
the same population of electrons, the steep X-ray spectrum implies
that the X-ray emission is in the tail of the spectral distribution of
the synchrotron radiation ($\nu> \nu_c$, the critical frequency).
Alternatively, Stevens et al. (1999) pointed out that the radio
feature may represent enhanced star formation often seen at the ends
of bars in barred galaxies. This would result in a soft X-ray spectrum
similar to the thermal emission from hot gas in the circumnuclear
ring, as we see here ($kT \sim 0.8\pm 0.04$ keV) if we fit the jet
spectrum with a single temperature thermal emission model.  We
examined the high angular resolution (0.07\arcsec) {\em HST}/FOC
F437M($\lambda_0=4290\AA$) image that was presented in Kristen et
al.(1997). None of the reported compact continuum source is associated
with the jet, with an observed flux completeness limit corresponds to
$B=19.3$ mag. No apparent mid-IR source is found at the location of
the jet in the {\em Spitzer} IRAC images.  The lacking in optical and
IR counterparts seems to disfavor the enhanced star formation origin
for the X-ray emission.  Neither the radio observation nor our X-ray
image show evidence for a counter jet. If indeed we are seeing a radio
jet, then the observed one-sidedness could be explained by
relativistic beaming with a jet velocity 0.7$c$, or with free-free
absorption by the circumnuclear disk as suggested by Hjelm \& Lindblad
(1996).  It is also interesting to note that the X-ray-to-6cm
luminosity ratio $L_X/L_{6cm}$ is $\sim 240$ for the jet feature (6 cm
luminosity of the jet from Stevens et al. 1999), similar to that of
radio SNe and radio quiet quasars (e.g., a few hundred, Stevens et
al. 1999; Landt et al. 2001).

\subsection{Comparison with Other Galaxies: AGN Feedback vs. Starburst Feedback}\label{comp2}

NGC 1365 is a LIRG ($L_{FIR}=10^{11}L_{\odot}$; Ghosh et al. 1993;
Sanders et al. 2003). The central nucleus contributes little (few
percent; Komossa \& Schulz 1998) to the total IR luminosity, most of
which could be attributed to the active star forming ring. Sakamoto et
al. (2007) found that starburst in subregions of the NGC 1365
circumnuclear ring alone is comparable to the entire starburst in
galaxies like M82 and NGC 253. The star formation rate in the inner
kpc region is $\Sigma_{SFR}=3$\MSUN yr$^{-1}$kpc$^{-2}$, among the
highest in the non-merging galaxies (Kennicutt 1998).

For star-forming galaxies, a good correlation between the X-ray and
the far infrared (FIR) luminosities has long been known and
extensively investigated (e.g., Fabbiano, Gioia \& Trinchieri\ 1988;
David et al. 1992; Fabbiano \& Shapley 2002; Ranalli et al. 2003;
Colbert et al. 2004; Persic \& Rephaeli 2007). The X-ray emission in
the circumnuclear starburst mainly comes from the young SNe (also
SNRs) and X-ray binaries (XRBs; Fabbiano 1989), the formation of which
requires time for the stellar evolution of O and B stars to complete.
Therefore X-ray emission may trace the evolution of a starburst with
some delay (a few Myr, the main-sequence life time of a massive star),
and the X-ray-to-FIR luminosity ratio $L_X/L_{FIR}$ could reflect the
age of the starburst event (see Swartz et al. 2006).  Recently
Mas-Hesse et al. (2008) computed soft X-ray and FIR luminosities for
different starburst scenario (instantaneous or extended), using
realistic efficiency of converting mechanical energy into soft X-ray
emission, and concluded that the soft X-ray emission can be used to
trace the star formation rate in relatively young and unevolved
starbursts.

In Table~\ref{starburst} we compiled $L_X$ (0.1--2.5 keV; total
line-of-sight absorption corrected to represent intrinsic luminosity),
IRAS FIR luminosity $L_{FIR}$, and measured ages of young clusters
associated with starburst events for a number of starburst galaxies
(Condon et al. 1998, Sanders et al. 2003; see complete reference list
in Table footnote). For NGC 1365, we calculated the $L_X/L_{FIR}$
using both total (non-nuclear) X-ray emission over total FIR emission,
and X-ray emission from circumnuclear starburst ring over FIR emission
from the ring. Given the available data, FIR emission from the NGC
1365 ring cannot be accurately measured due to the low spatial
resolution at 70$\mu$m and 160$\mu$m (saturated) {\em Spitzer} images.
To derive an approximate $L_{FIR}$ value for the ring, we estimate the
$L_{IR}$ ratio between the nucleus and the nucleus combined with the
ring at 24$\mu$m, which is $\sim$50\%. Together the nucleus and the
ring contribute $\sim$70\% of the FIR emission from NGC 1365, therefore
we estimate that $L_{FIR}$ of the ring is approximately 35\% of the
IRAS $L_{FIR}$ for NGC 1365.  The $L_X/L_{FIR}\sim 10^{-4}$ is in good
agreement with that of M82, whose $L_X/L_{FIR}$ represents a sample of
starburst galaxies (Strickland et al. 2004).  In comparison, the
ultra-luminous IR galaxy (ULIRG) Arp 220 has a lower $L_X/L_{FIR}\sim
3\times 10^{-5}$ (Soifer 1984; McDowell et al. 2003), which hosts very
young clusters (1--3 Myr old; Wilson et al. 2006). On the other hand,
the merging system Antennae shows a higher $L_X/L_{FIR}$, where older
clusters (age $\sim$13 Myr) have been found (Mengel et al. 2001).
Another well known starburst galaxy NGC 253 had a starburst event
20--30 Myr ago (Engelbracht et al. 1998). Its $L_X/L_{FIR}\sim
10^{-4}$ is a lower limit because of the high obscuration of soft
X-ray emission in the starburst nucleus, which could be similar to
that of the Antennae.  We also included the Galactic center, if viewed
from outside of our Galaxy, the Galactic center will appear as a
circumnuclear starburst with a weakly active nucleus (Sgr A$^{\ast}$).

Figure~\ref{LxLFIR} shows the resulting $L_X$ vs. $L_{FIR}$ plot and
the predicted values for 1 Myr, 4 Myr, and 6 Myr old starbursts from
Mas-Hesse et al. (2008). The different assumptions on star formation
history and energy conversion efficiency yield similar predictions for
such relatively unevolved systems.  Their locations relative to the
model tracks show that the starburst in NGC 3351 is over 1 Myr
old. NGC 1365, M82, and NGC 253 lie closely with the 4 Myr model
tracks, and the older merging system Antennae shows values consistent
with the 6 Myr model.  Comparing to the ages obtained in the
literature (see Table~\ref{starburst}), these values derived from the
model tracks appear to be qualitatively consistent but a few Myrs
younger.  Many factors may contribute to this: the assumption of the
star formation history in the galaxies is apparently over-simplified;
the predicted age in the model is coupled with the efficiency of
re-processing mechanical energy into soft X-ray emission, which could
vary in different ISM environment; the measured values from literature
also have large range as charting the age of starbursts is challenging
(see Gallagher \& Smith 2005 review); the measured soft band $L_X$ can
be significantly underestimated because of the heavily obscured
starburst regions (see related discussion in Table~\ref{starburst}
footnotes). This raises caution that the comparison with theoretical
evolution may only be used to roughly estimate the age of starbursts.

It is known that the emergent soft X-ray luminosities may be greatly
reduced from the star forming galaxies with significant visual
extinction\footnote{Combining $A_V\approx 4.5\times 10^{-22} N_H$
  (mag) (Ryter 1996) with the X-ray absorption cross sections from
  Morrison \& McCammon (1983) gives $\tau_X \ga 1$ at $E\approx 2$ keV
  when $A_V\ga 10$.}.  However, accurate attenuation corrections are
unrealistic, the intrinsic absorption columns towards the galaxies are
inhomogeneous and difficult to measure. For example, the extinction to
the core of Arp 220 is estimated at $A_V > 100$ mag from sub-mm
continuum measurements (Sakamoto et al. 2008), while the revealed star
clusters have $A_V\sim 3$ mag (Wilson et al.2006).  To correct for
absorption, we adopt $N_H$ from X-ray spectral fit if available,
otherwise we convert published $A_V$ that are typical of the star
clusters to $N_H$ using the Galactic gas-to-dust ratio.  For the
galaxies with $\log N_H\sim 21$, the uncertainty in the absorption
columns results in a $\pm$0.08 dex uncertainty in the absorption
corrected $L_{X,corr}$ from the XSPEC spectral fit.  For Arp 220 and
NGC 253, correcting for a $\log N_H\sim 23$ column (the most obscured
nuclear region) can increase $L_{X,corr}$ by a factor of 100 (see Lutz
et al. 2004 for estimated correction in type 2 Seyferts).  In
comparison, the correlation between the X-ray luminosities (0.5--4.5
keV band) and the FIR luminosities derived in David et al. (1992)
assumes an average $N_H=3\times 10^{20}$ cm$^{-2}$ for all FIR bright
galaxies; the empirical calibration between the soft X-ray
luminosities (0.5--2.0 keV) and the FIR luminosities in Ranalli et
al. (2003) is corrected for Galactic absorption only.

Swartz et al. (2006) reported {\em Chandra} observations of the nearby
($D=10$ Mpc) barred spiral galaxy, NGC 3351, which has a prominent kpc
wide circumnuclear ring in reminiscent of the NGC 1365 ring. The X-ray
emission in the NGC 3351 ring has $kT\sim 0.5$ keV ($L_X\sim 6.5\times
10^{38}$ ergs s$^{-1}$, 0.3--3.0 keV), attributed to hot diffuse gas
associated with starburst.  The X-ray luminosity of the central
nucleus is estimated to be $Lx<10^{37}$ ergs s$^{-1}$ (0.5--8.0 keV),
and X-ray photoionization is considered unimportant.  The extended
soft X-ray emission, similar to what we see in NGC 1365, is
interpreted as gas outflow from the star-forming ring out of the
galactic plane.

A multiwavelength study of the circumnuclear extended emission in
Seyfert 2 galaxy NGC 2110 (Evans et al. 2006) suggested that shock-heated
multiphase gas can account for the observations, although photoionized
gas by the nucleus cannot be ruled out. In contrast, in another nearby
AGN-starburst galaxy NGC 6764, Croston et al. (2008) detected
extended, $kT=0.75$ keV X-ray emission coincident with the radio
bubbles, likely inflated by the AGN (Hota \& Saikia 2006).  They rule
out a galactic wind as the origin of the X-ray emission and instead
favor hot gas heated by jet/ISM interaction in the AGN outflow,
similar to the radio lobes in radio galaxies (e.g., Centaurus A, Kraft
et al. 2003).  This is quite a different situation from NGC 1365,
which shows a weak radio nucleus and jet, and no radio bubble
(Sandqvist et al. 1995).

Comparison of NGC 1365 to these starburst and Seyfert 2 galaxies
suggests that, in galaxies where the AGN is not the dominant energetic
factor in the circumnuclear environment (i.e., where a starburst plays
major role), the extended soft X-ray emission shows a thermal origin
as SNe and massive stars drive outflow into the disk and halo
(galactic winds, see Heckman et al. 1990, Strickland et al. 2004).  If
instead the energetics of nuclear radiation suffice, circumnuclear
X-ray emission may be solely attributed to AGN photoionization (e.g.,
Ogle et al. 2003, Bianchi et al. 2006).  When a powerful AGN
jet/outflow is seen as evidenced by large scale radio bubbles, the
interaction with the ISM may heat the entrained gas and account for
the X-ray emission (e.g., Kraft et al. 2003, Croston et al. 2008).
These galaxy feedback processes are critical in our understanding of
galaxy formation and evolution, which can be best studied when high
resolution radio, X-ray, and optical NLR data are combined (e.g.,
Evans et al. 2006; Bianchi et al. 2007).

\section{Summary}

We have presented a detailed {\em Chandra}/ACIS imaging study of the
nearby Seyfert galaxy NGC 1365. Based on the high resolution 100 ks
images, we confirm results from previous X-ray studies and present new
findings in the complex circumnuclear region.
 
We have created an X-ray broad band color image as well as narrow band
line-emission image, showing a biconical soft X-ray emission region
extending from the nucleus, coincident with the high excitation
outflow cones seen in [OIII]/H$\alpha$ emission line images. A
prominent kpc-scale circumnuclear star-forming ring is seen emitting
harder X-rays.  The spectral fitting on the spatially separated
components indicates thermal plasma origin for the X-ray emission in
the ionization cones, which is also supported by the spatial
anti-correlation between X-ray emission and the [OIII] emission.  The
abundance ratios for regions of NGC 1365 are fully consistent with the
theoretical values for enrichment from Type II SNe.  X-ray morphology
of NGC 1365 is compared with {\em Spitzer} mid-IR images and [OIII]
emission line observations.  We argue that pure photoionization by AGN
cannot account for the extended soft X-ray emission based on the
spectral fit, energetic estimate, and morphology evidences. We
attribute the observed soft X-rays to thermal emission from shock
heated hot gas, although we cannot rule out a photoionized
contribution in the line emission.  X-ray emission from a previously
reported nuclear radio jet is detected.

Further progress will require deeper {\em Chandra} images of NGC 1365,
which we plan to obtain in the future to enable studies of the
morphology and spectral properties in greater details (similar to
Baldi et al. 2006a,b), with better statistics and constraints.
Isolation of discrete sources in the deep images will allow removal of
contribution from the XRB population to the diffuse emission. This
enables investigations to distinguish the nature of X-ray hot-spots
from the truly diffuse emission they embedded in, and studies of the
X-ray luminosity function of XRBs in NGC 1365.  Spatial complexity of
the X-ray emission can be overcome by dividing the circumnuclear ring
and diffuse cones to smaller regions.  This will also localize any
abundance anomalies. With more counts in the diffuse emission, we will
be able to perform spectral modeling on the radial gradient with
different models and search for variation in the X-ray structure and
spectra.  Expecting that a photoionized wind will have constant
ionization parameter, and that a thermal wind will cool radially, we
plan to use this diagnostic to unambiguously test the photoionized
wind model and thermal wind model for the origin of X-ray emission
associated with the ENLR in NGC 1365. Radially constant ionization
parameters have been found in a sample of type 2 Seyferts (Bianchi et
al. 2006), suggesting that at least in some cases, soft X-ray and
optical NLR come from a single photoionized medium.  Deeper X-ray
imaging will also clarify the nature of the jet-like feature.

\acknowledgements

We thank an anonymous referee for helpful comments that improved the
clarity of the manuscript.  We are grateful to S. Veilleux for
providing us the [OIII] images in Veilleux et al. (2003), and Tim
Kallman for the helpful information about XSTAR. J. W. thanks A. Baldi
for his help in the data analysis, H. Oti Floranes for providing the
model results in Mas-Hesse et al. (2008), and A. Siemiginowska for
helpful discussion on radio jets. G. F. and M. E. acknowledge
stimulating discussions at the Aspen Center for Physics workshop on
Active Galactic Nuclei. This work is supported by NASA Contract
NAS8-39073 (CXC) and {\em Chandra} GO Grant G06-7102X (PI: Risaliti).


\clearpage

\begin{figure}[H]
\epsscale{0.4}
\plotone{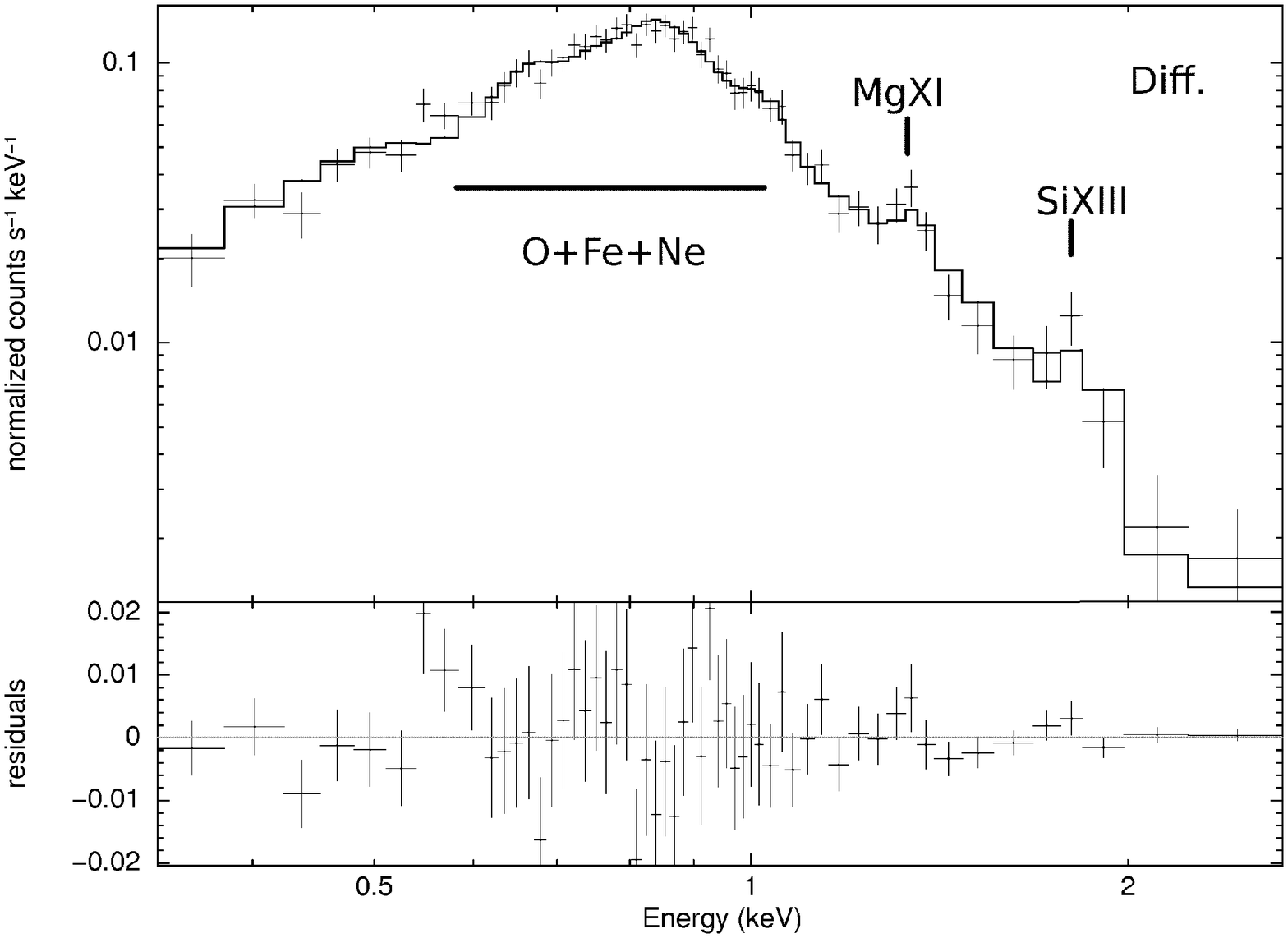}
\plotone{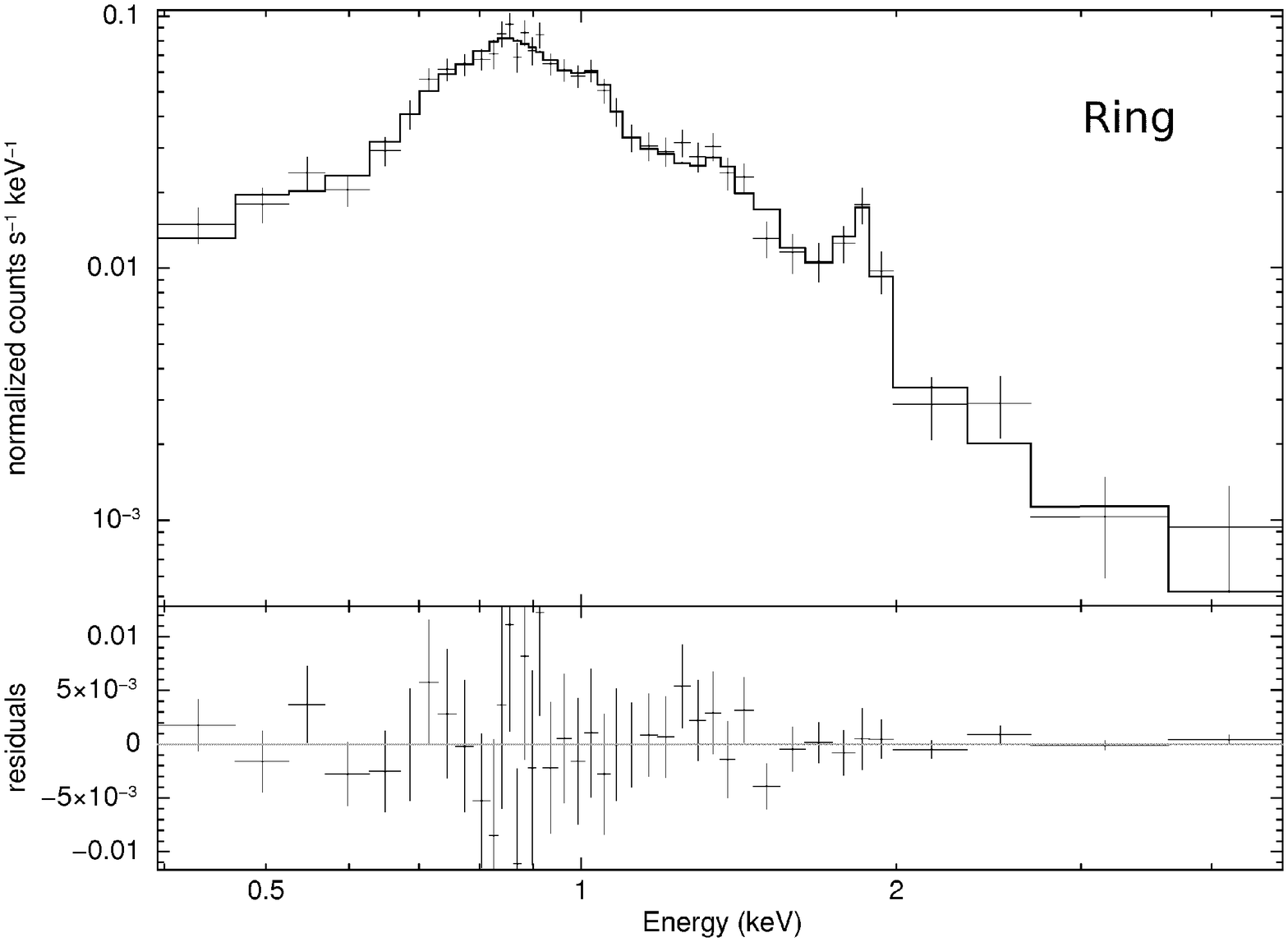}
\plotone{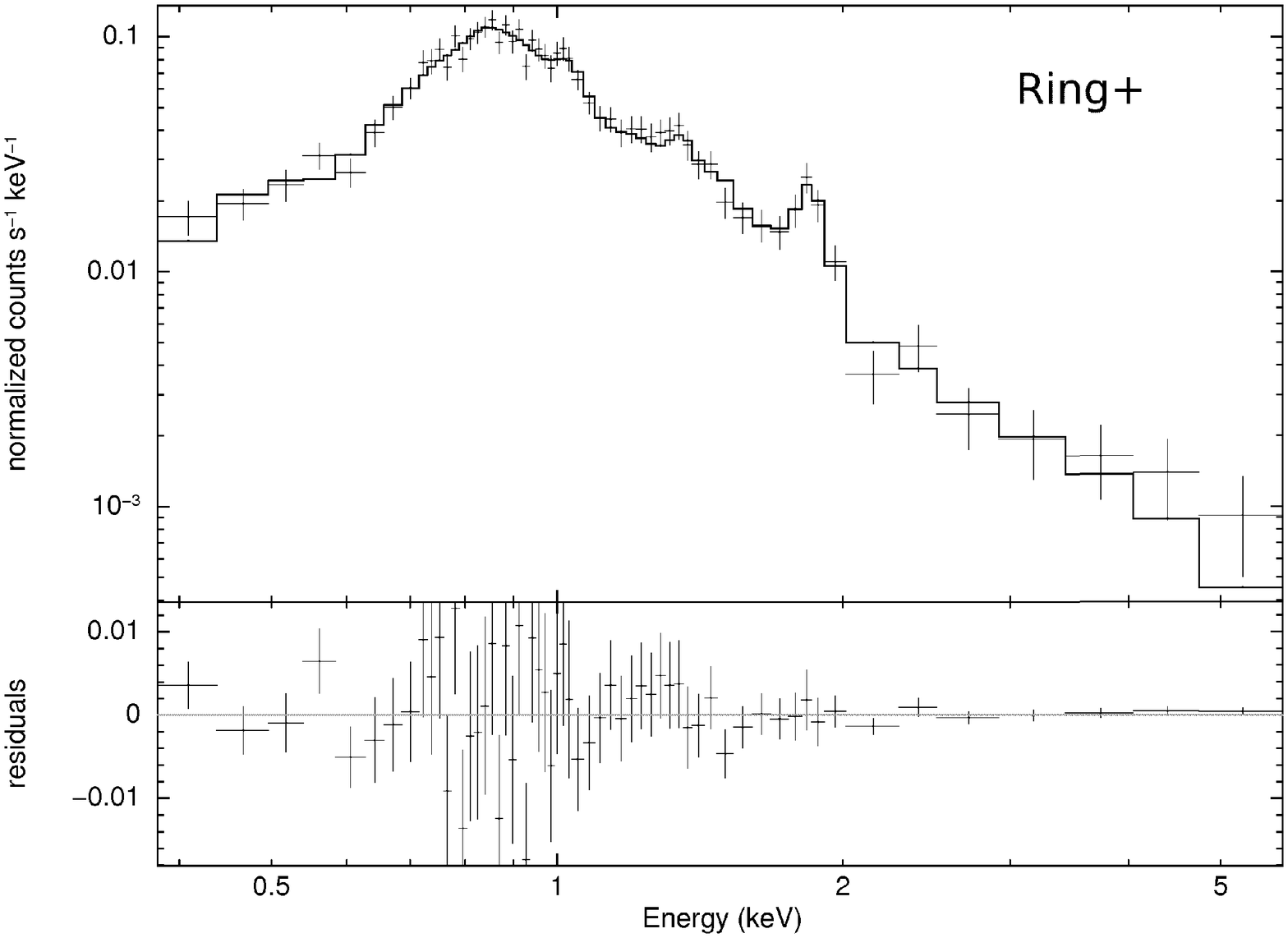}
\plotone{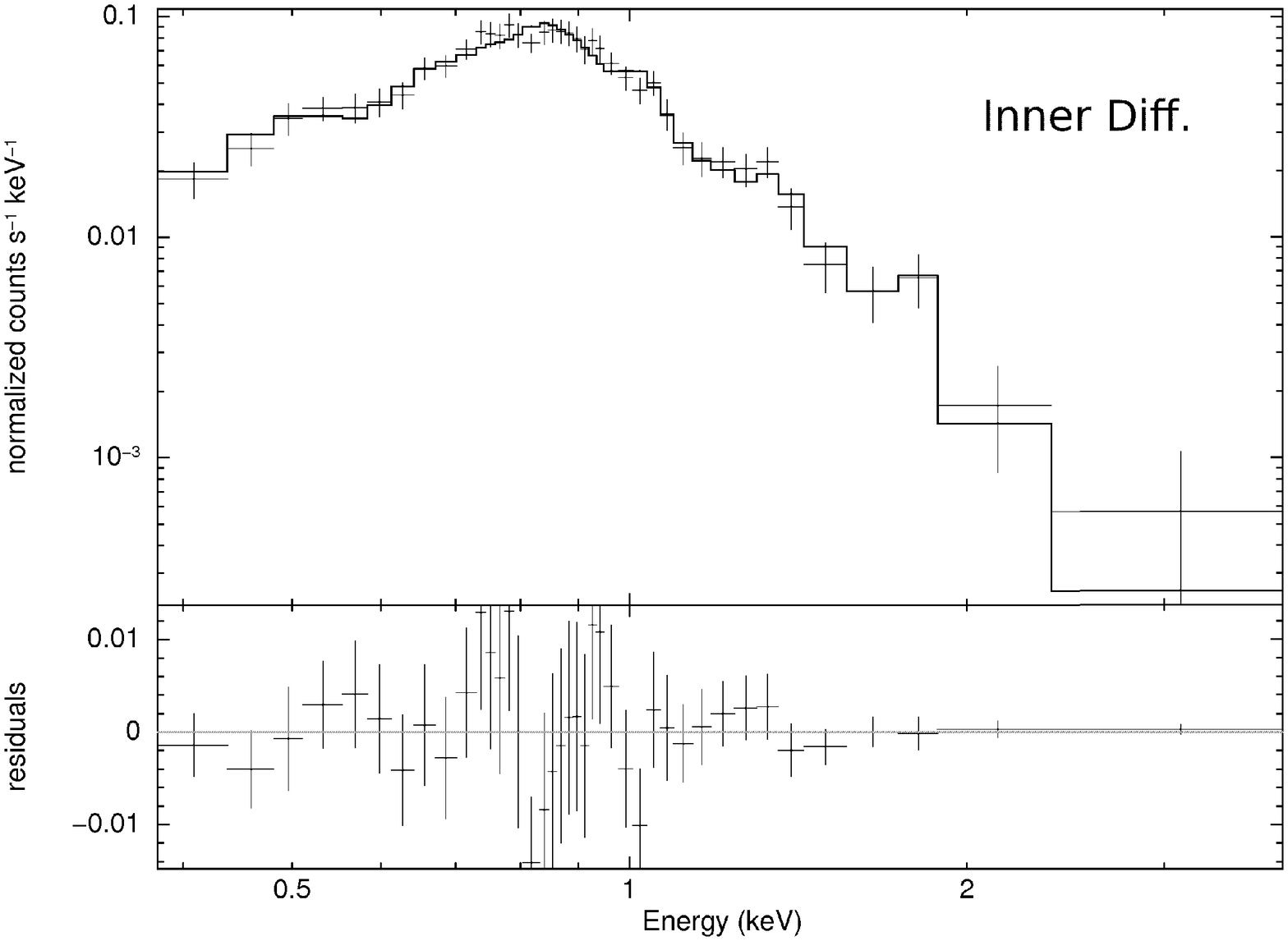}
\caption{X-ray spectra and best-fits for the regions listed in
  Table~\ref{mytable1}.
\label{spectra}}
\end{figure}

\addtocounter{figure}{-1}
\begin{figure}[H]
\epsscale{0.4}
\plotone{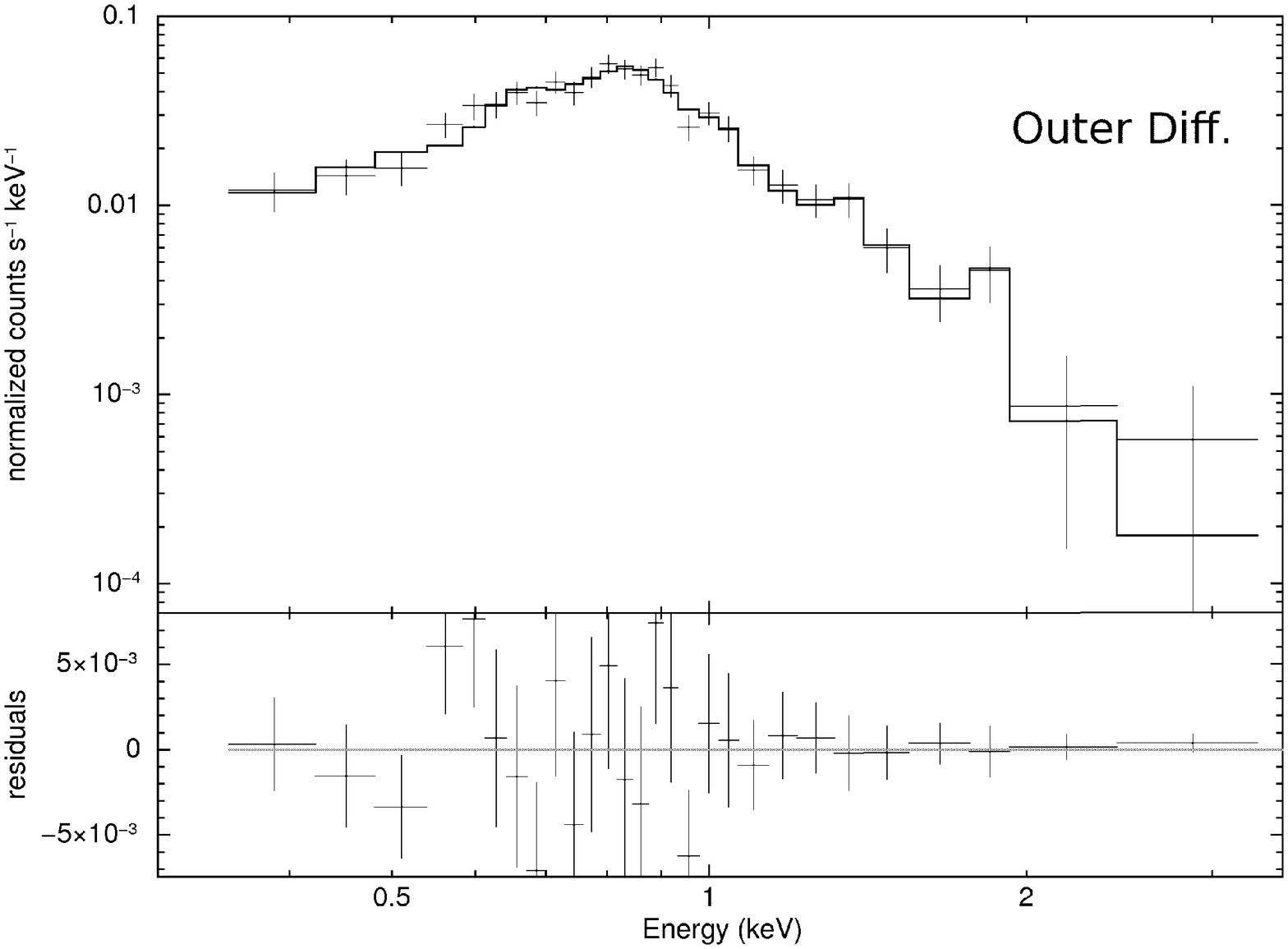}
\plotone{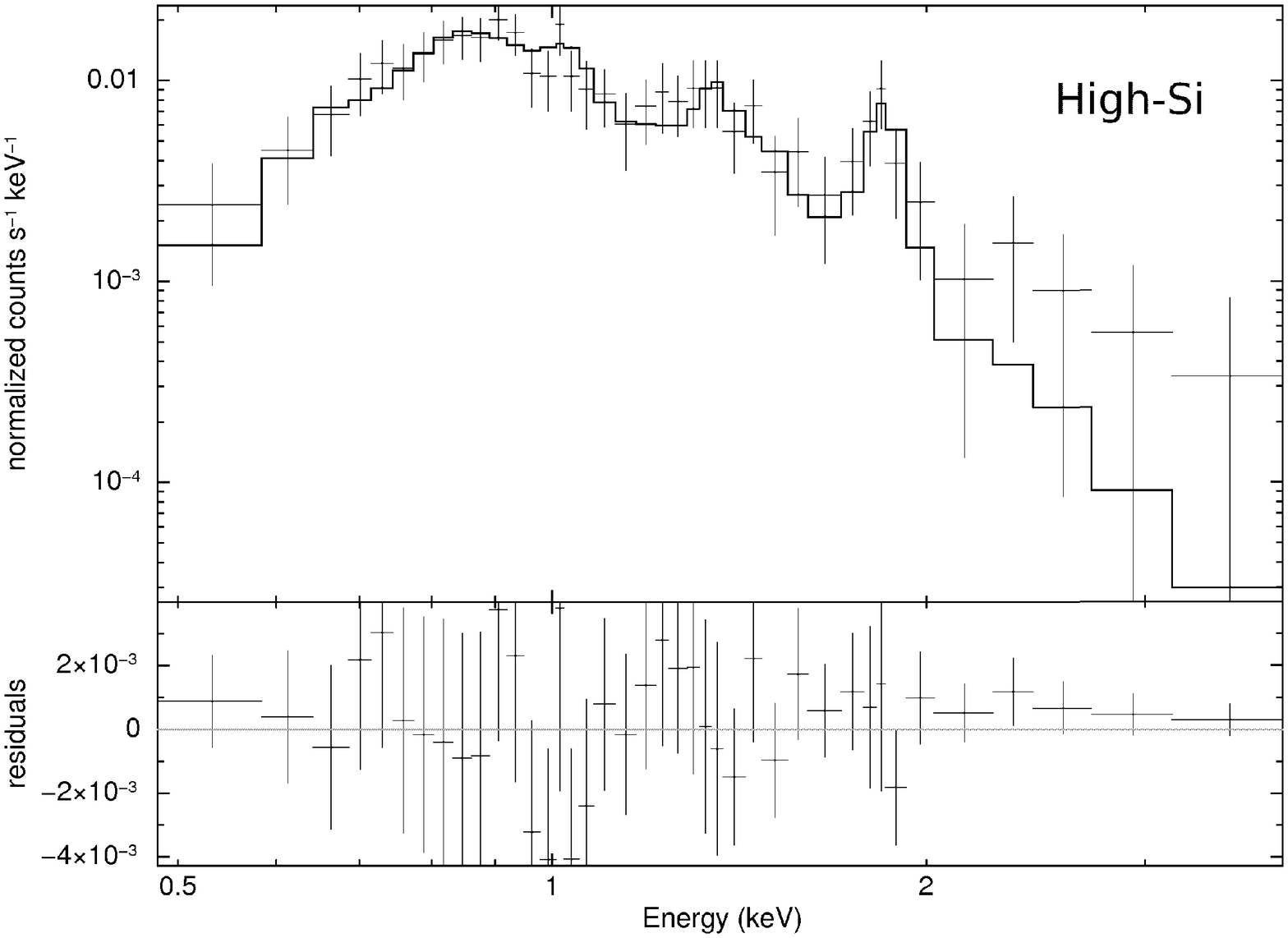}
\plotone{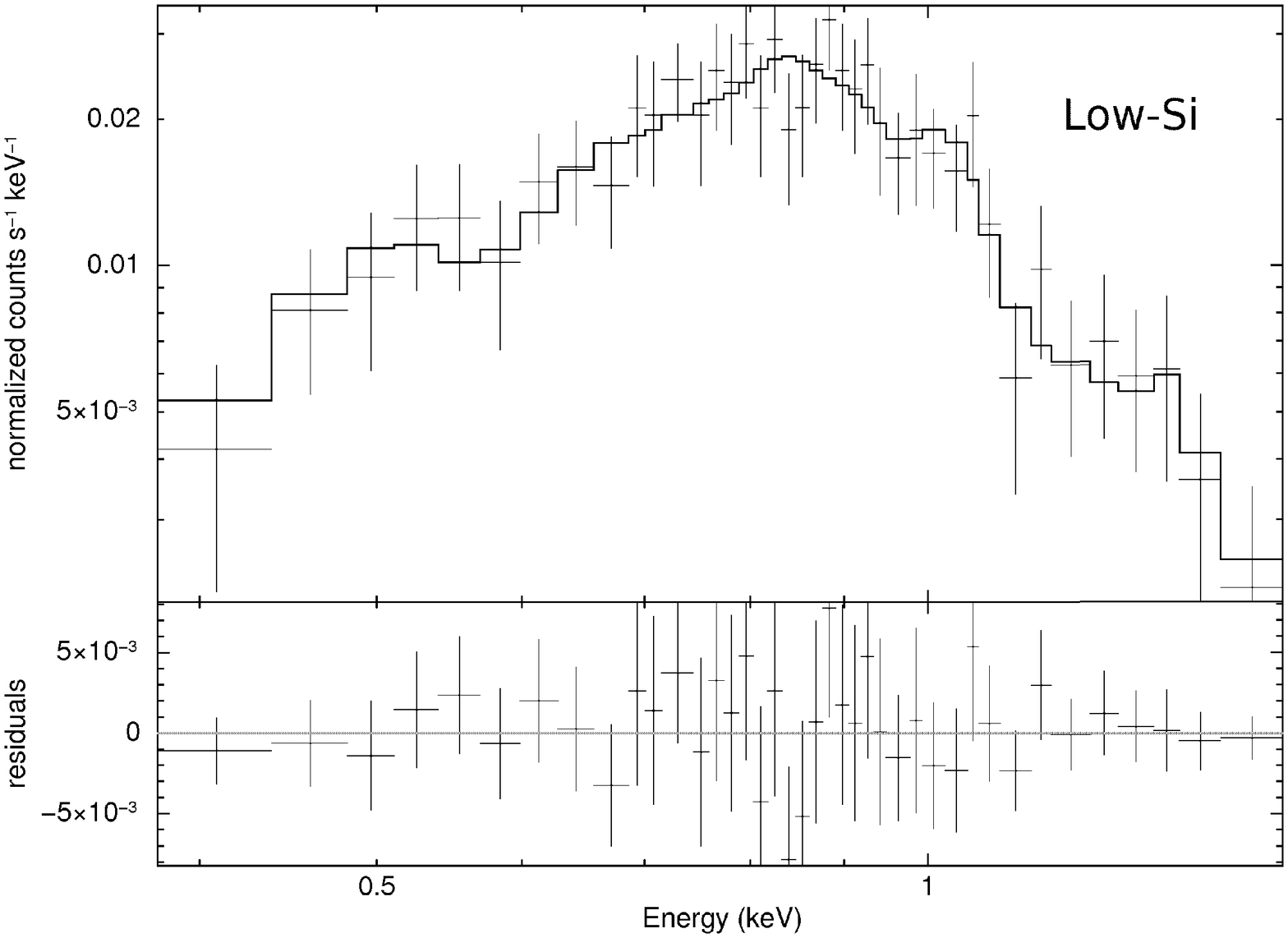}
\plotone{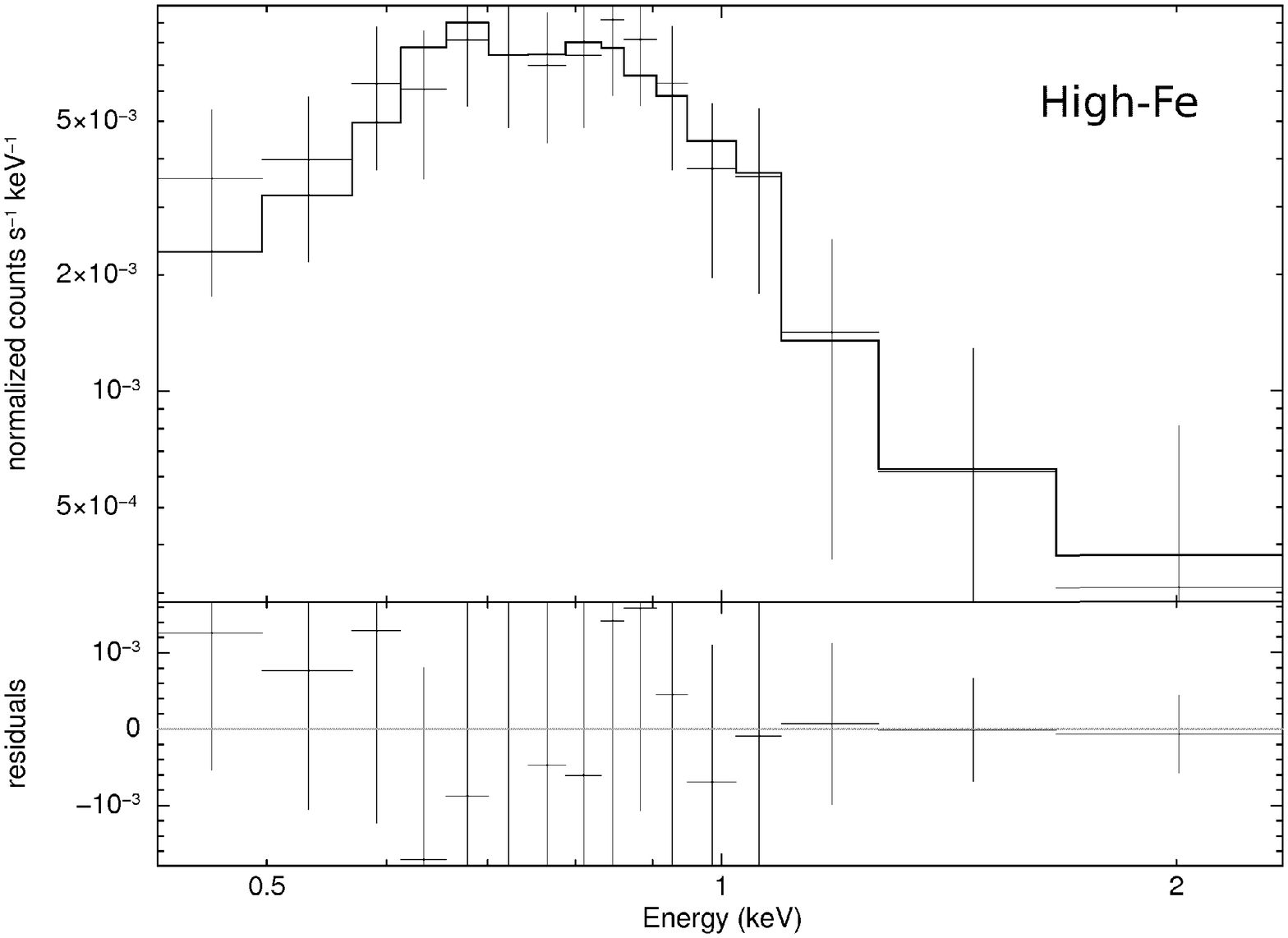}
\caption{Continued. X-ray spectra and best-fits for the regions listed in
  Table~\ref{mytable1}.
}
\end{figure}

\begin{figure}[H]
\epsscale{0.45}
\plotone{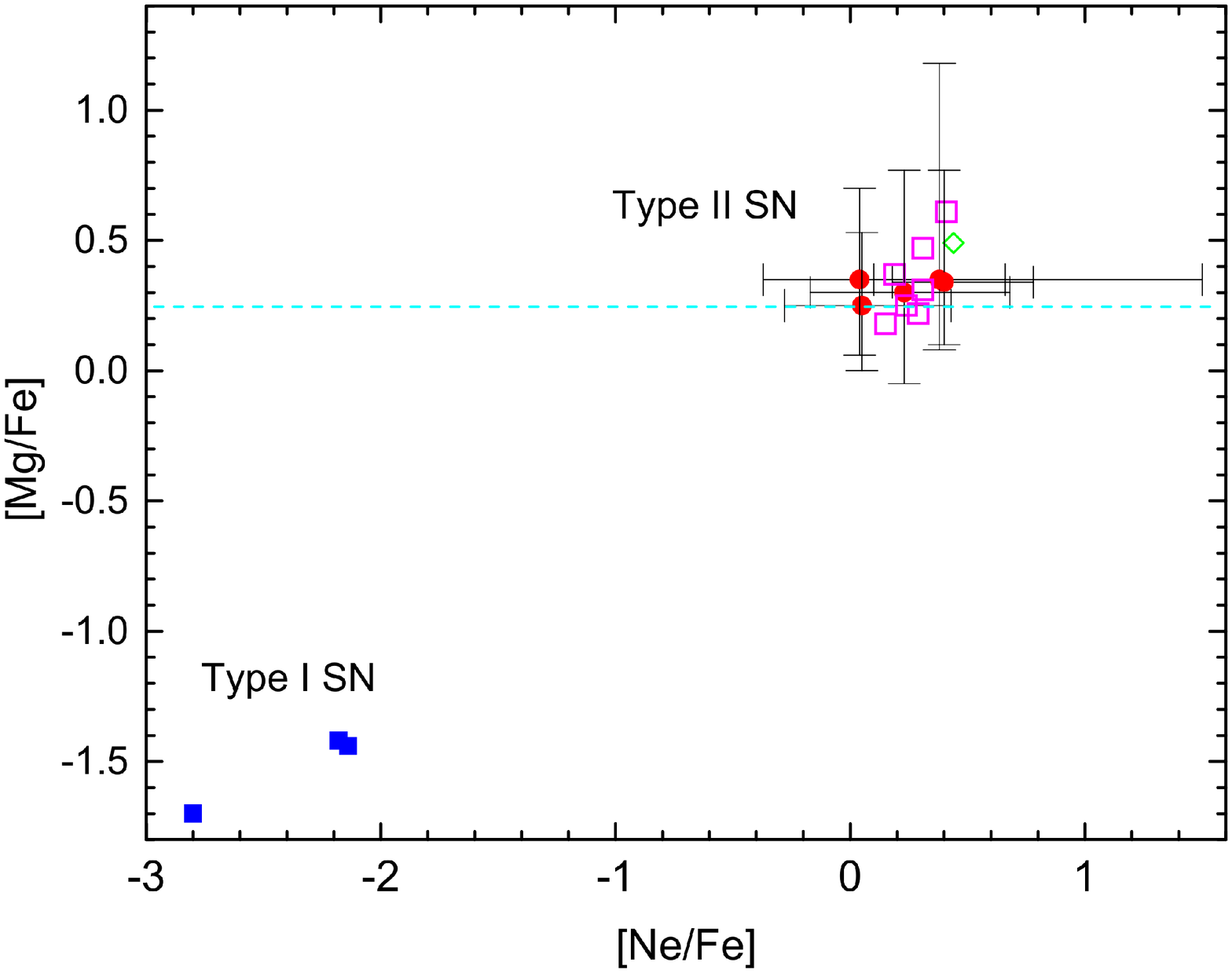}
\plotone{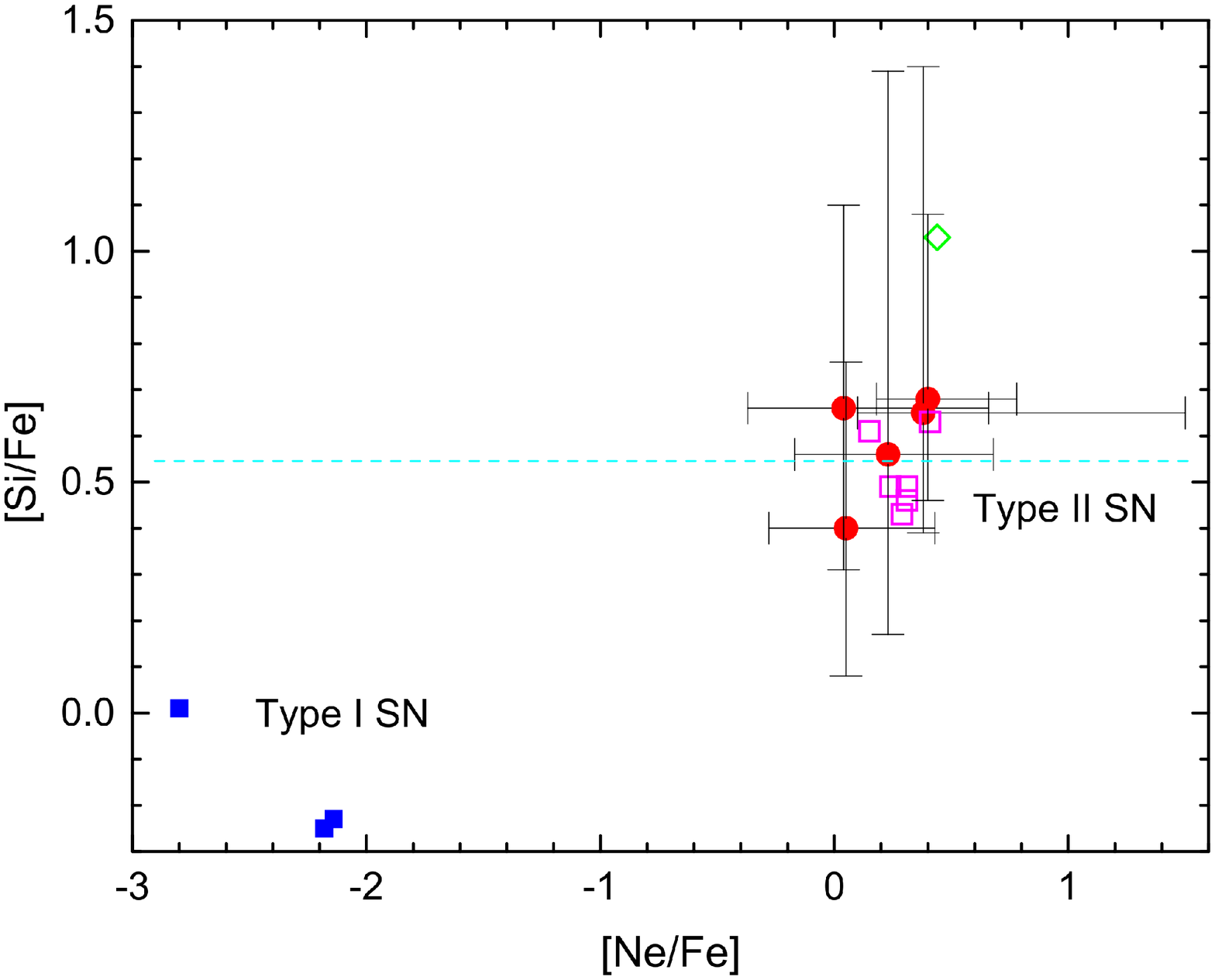}
\plotone{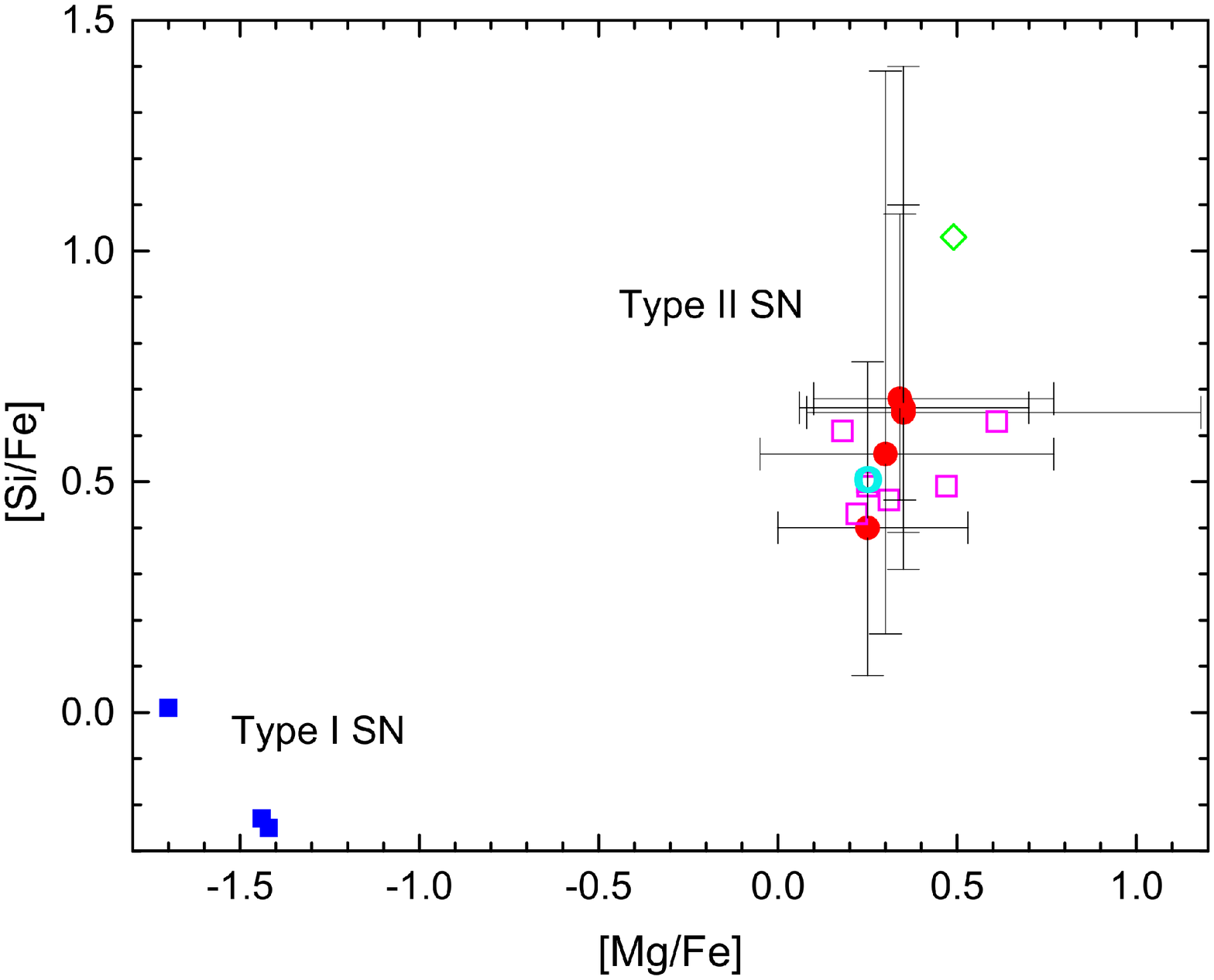}
\caption{Abundace ratio diagrams for regions in NGC 1365: [Ne/Fe]
  vs. [Mg/Fe], [Ne/Fe] vs. [Si/Fe], and [Mg/Fe] vs. [Si/Fe]. Abundance
  units refer to meteoritic abundances of Anders \& Grevesse (1989).
  The regions in Table~\ref{mytable2} are plotted with filled circles
  (red). The ratios for stellar yields from SNe Type I (blue filled
  squares) and Type II (magenta open squares) are also shown, adopted
  from various theoretical works (see Table~\ref{mytable3} for the
  list of references). Diamond (green; upper limit for [Si/Fe]) and
  open circle (cyan; line is used where no [Ne/Fe] measurement is
  available) denotes abundance ratio for Strickland et al.(2004)
  sample of starbursts and the warm Galactic halo, respectively.
\label{NeMg}}
\end{figure}

\begin{figure}[H]
\epsscale{0.9}
\plotone{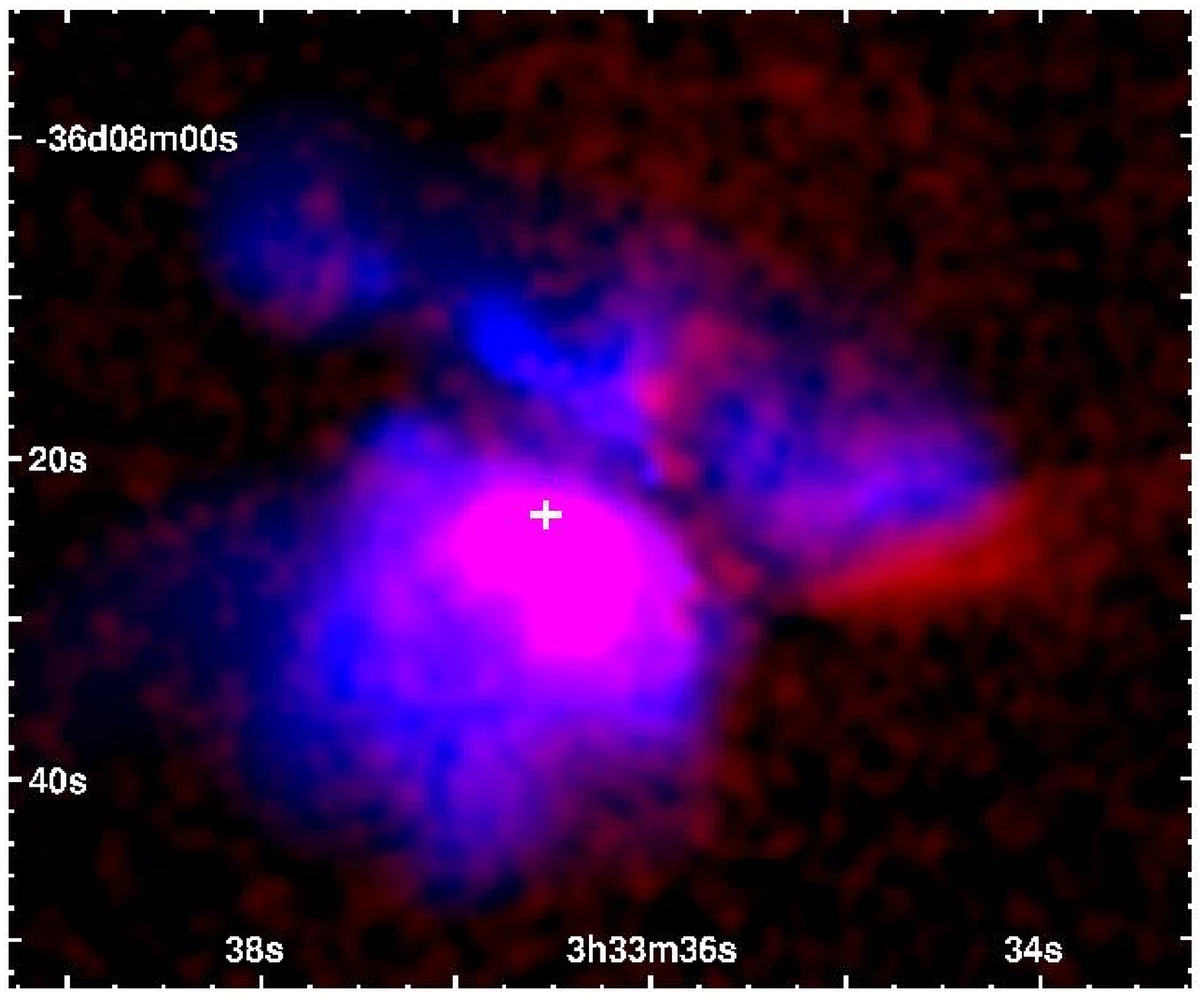}
\caption{The composite image of the soft band (0.3--0.65 keV) X-ray
  emission (blue) and continuum subtracted [OIII] image (red).  The
  plus symbol marks the position of the nucleus. Note how the X-ray
  morphology appears to complement the [OIII] emission in the NW
  region.  The [OIII] image is from Veilleux et al. (2003).
\label{oiii}}
\end{figure}

\begin{figure}[H]
\epsscale{0.6}
\plotone{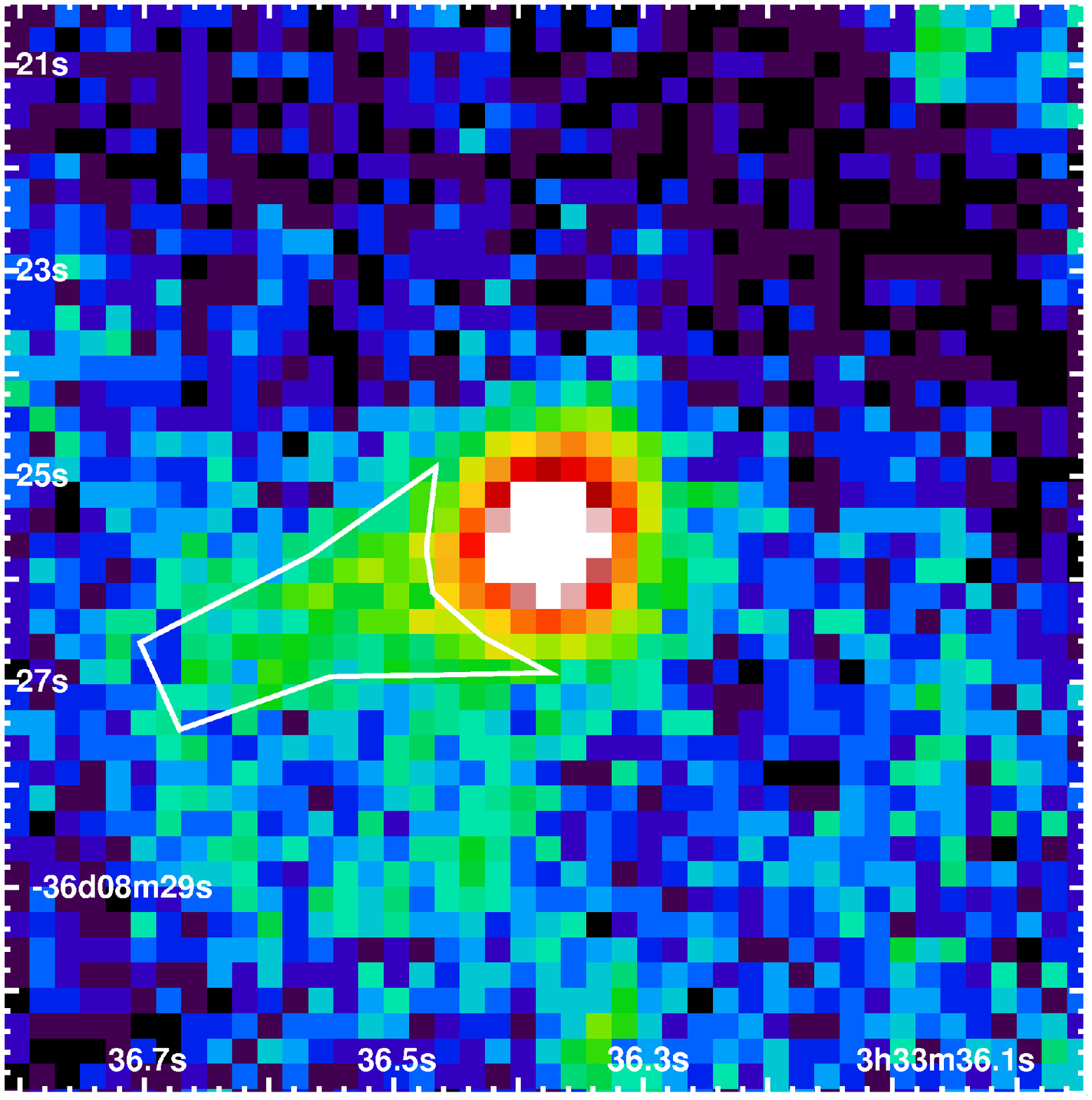}
\caption{The $10\arcsec\times 10\arcsec$ region around the nucleus in
  the full band (0.3--7 keV) image ($0.25\arcsec$/pixel binning). A
  4$\arcsec$ elongation is seen. A spectrum for the jet-like feature
  is extracted from the region shown by the white polygon.
\label{jet}}
\end{figure}

\begin{figure}[H]
\epsscale{0.9}
\plotone{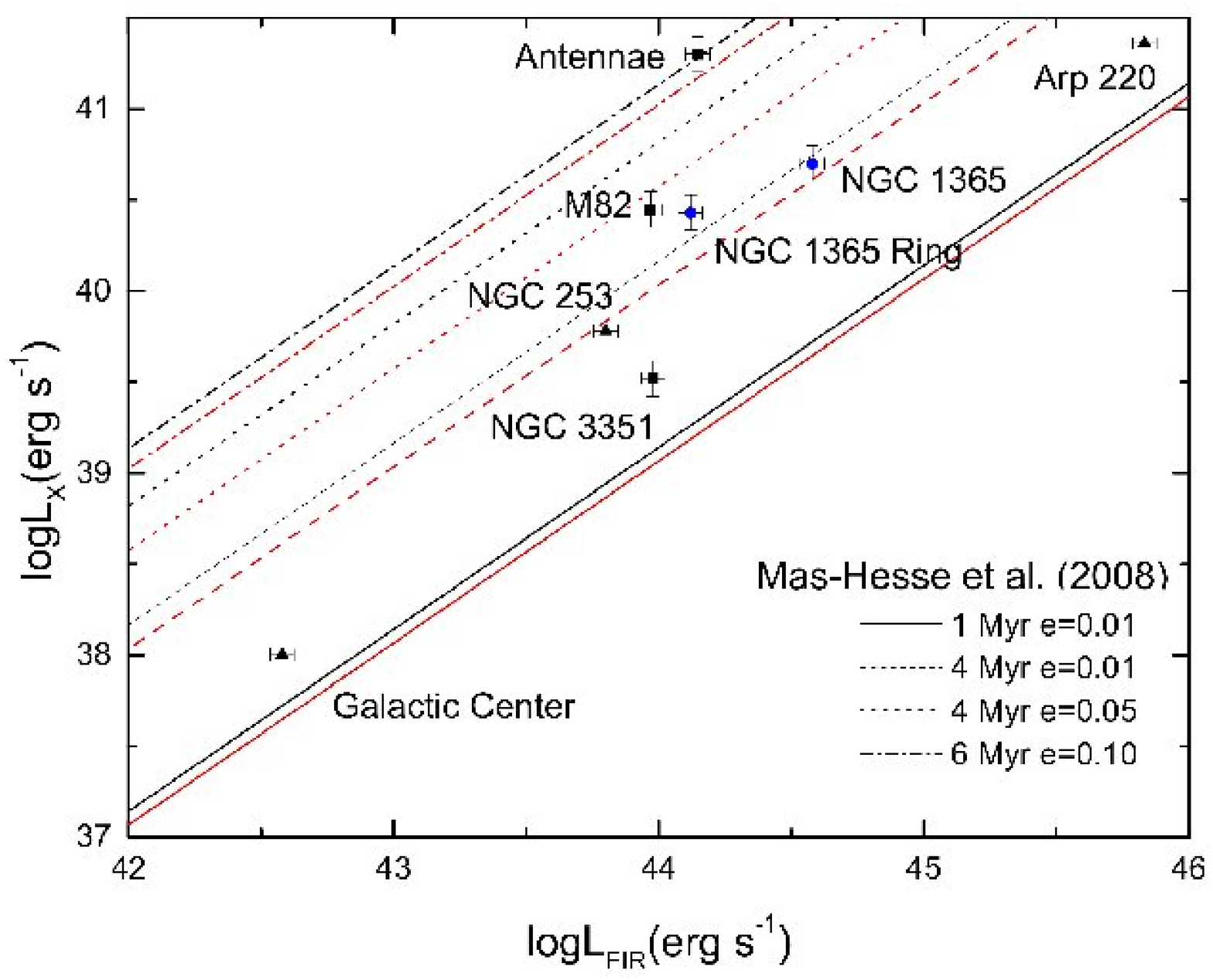}
\caption{Comparison of the X-ray and FIR luminosities of NGC 1365 and six other galaxies. The predictions of soft X-ray and FIR luminosities from evolutionary synthesis models in Mas-Hesse et al. (2008) are shown as lines (see legend) for different starburst ages, $e$ values (the efficiency of re-processing mechanical energy into soft X-ray emission, and star formation history: instantaneous starburst (see Mas-Hesse et al. 2008 for detailed model description; all black lines) vs. extended starburst (all red lines). The absorption corrected $L_X$ for NGC 253, Arp 220, and the Galactic Center is considered as lower limit because of the high intrinsic obscuration in these regions. See also Table~\ref{starburst}.
\label{LxLFIR}}
\end{figure}

\clearpage


\begin{deluxetable}{llcccc}
\tabletypesize{\small} \tablecaption{Log of {\it Chandra} ACIS-S
Observations of NGC 1365\label{obslog}} \tablewidth{0pt}
\tablehead{\colhead{Obs ID} & \colhead{Date} & \colhead{Pointing R.A.}
& \colhead{Decl.} & \colhead{Roll Angle ($^\circ$)} & \colhead{Good time
interval (ks)}} \startdata

3554&December 24, 2002& 53.409638& -36.147430& 321.9&13.0\\
6868&April 17, 2006& 53.399219& -36.139588& 215.0&14.6\\
6869&April 20, 2006& 53.399299& -36.139509&211.8&15.5\\
6870&April 23, 2006& 53.399377& -36.139370&208.1&14.6\\
6871&April 10, 2006& 53.399118& -36.139908&223.9&13.4\\
6872&April 12, 2006& 53.399126& -36.139825&221.3&14.6\\
6873&April 15, 2006& 53.399201& -36.139764&218.4&14.6\\
\enddata
\tablenotetext{(1)}{ObsID 3554 is done with full ACIS-S array, while
the rest observations were done in 1/4 sub-array mode. All observations are performed in ``Timed Event, Faint'' mode.}

\end{deluxetable}


\clearpage
\pagestyle{empty}
\begin{deluxetable}{llcccccccccc}
\tabletypesize{\scriptsize} \rotate \tablecaption{Best-Fit
Parameters (1.65$\sigma$ errors for one interesting parameter) for
the Diffuse Emission of NGC 1365. \label{mytable1}}
\tablewidth{0pt} \tablehead{ \colhead{Region} &
\colhead{Net Cnts} &\colhead{$\chi^2/dof$} & \colhead{\begin{tabular}{c}
$N_H$\\
($\times10^{20}$ cm$^{-2}$)
\end{tabular}} &
\colhead{\begin{tabular}{c}
$kT$\\
(keV)
\end{tabular}}&
\colhead{\begin{tabular}{c}
$Z_{O}$\\
($\times Z_{O,\odot}$)
\end{tabular}} &
\colhead{\begin{tabular}{c}
$Z_{Ne}$\\
($\times Z_{Ne,\odot}$)
\end{tabular}} &
\colhead{\begin{tabular}{c}
$Z_{Mg}$\\
($\times Z_{Mg,\odot}$)
\end{tabular}} &
\colhead{\begin{tabular}{c}
$Z_{Si}$\\
($\times Z_{Si,\odot}$)
\end{tabular}} &
\colhead{\begin{tabular}{c}
$Z_{Fe}$\\
($\times Z_{Fe,\odot}$)
\end{tabular}} &
\colhead{$\Gamma$} &
\colhead{Mod\tablenotemark{(1)}}
}
\startdata
Diff.& $8216\pm224$&44.7/50&$3.59_{-2.60}^{+4.71}$&$0.57_{-0.03}^{+0.05}$& $0.38_{-0.21}^{+0.17}$&$0.20_{-0.15}^{+0.17}$&$0.32_{-0.17}^{+0.20}$&$0.45_{-0.32}^{+0.37}$&$0.18\pm0.04$&\nodata&1T\\

Ring&$5541\pm184$&21.8/31&$15.95_{-9.65}^{+12.08}$&$0.61\pm0.05$& 1.0& $7.73_{-3.37}^{+3.59}$&$6.65_{-3.33}^{+4.27}$&$14.74_{-6.26}^{+7.72}$&$3.06_{-0.83}^{+2.31}$&$2.65_{-0.49}^{+0.72}$&TP\\

Ring+&$7524\pm191$&34.7/47&$11.29_{-6.00}^{+6.63}$&$0.61_{-0.04}^{+0.03}$& 1.0&$3.39_{-1.85}^{+8.71}$&$3.17_{-1.61}^{+5.99}$&$6.23_{-2.95}^{+10.6}$&$1.41_{-0.51}^{+0.53}$&$2.35_{-0.42}^{+0.48}$&TP\\

Inner Diff.& $5330\pm242$&22/32&$8.51_{-3.83}^{+18.6}$&$0.48_{-0.17}^{+0.10}$&$0.14_{-0.07}^{+0.12}$&$0.17_{-0.15}^{+0.17}$&$0.20_{-0.15}^{+0.21}$&$0.36_{-0.31}^{+0.69}$&$0.10\pm0.03$&\nodata&1T\\

Outer Diff.&$3239\pm198$&19.8/25&$1.34*$&$0.57_{-0.06}^{+0.05}$& $0.58_{-0.24}^{+0.37}$&$0.24_{-0.22}^{+0.34}$&$0.49_{-0.31}^{+0.37}$&$1.02_{-0.79}^{+0.99}$&$0.22_{-0.05}^{+0.07}$&\nodata&1T\\

High-Si&$1382\pm165$&18/38&$33.14_{-23.77}^{+18.39}$&$0.61_{-0.10}^{+0.12}$& 1.0&1.0&$1.62_{-0.86}^{+2.29}$&$3.46_{-1.82}^{+3.95}$&$0.37_{-0.16}^{+0.29}$&\nodata&1T\\

Low-Si&$1580\pm166$&19/39&$1.34*$&$0.61\pm0.06$& 1.0& 1.0&$0.69_{-0.65}^{+0.74}$&$0$&$0.30_{-0.09}^{+0.12}$&\nodata&1T\\

High-Fe&$498\pm160$&10/32&$1.34*$&$0.46\pm0.09$& 1.0&1.0&$0.67_{-0.64}^{+3.77}$&$1.91_{-1.91}^{+51}$&$0.32_{-0.30}^{+0.97}$&\nodata&1T\\

\enddata

\tablenotetext{(1)}{* indicates freezed parameter. XSPEC Models: 1T = {\em tbabs(vapec)}; TP = {\em tbabs(vapec+powerlaw)}}

\end{deluxetable}
\clearpage


\begin{deluxetable}{ccccccc}
\rotate
\tabletypesize{\footnotesize}
\tablecaption{Emission Parameters for the Diffuse Emission in NGC 1365\label{lumin}}
\tablewidth{0pt}
\tablehead{
\colhead{\begin{tabular}{c}
Region \#\\
(1)
\end{tabular}} &
\colhead{\begin{tabular}{c}
$F_{0.3-10\:keV}^{p.l.}$\\
($\times10^{-14}$ erg cm$^{-2}$ s$^{-1}$)\\
(2)
\end{tabular}} &
\colhead{\begin{tabular}{c}
$F_{0.3-10\:keV}^{therm}$\\
($\times10^{-14}$ erg cm$^{-2}$ s$^{-1}$)\\
(3)
\end{tabular}} &
\colhead{\begin{tabular}{c}
$EM$\tablenotemark{(*)}\\
($\times10^{62}$ cm$^{-3}$)\\
(4)
\end{tabular}} &
\colhead{\begin{tabular}{c}
$L_{0.3-10\:keV}^{p.l.}$\\
($\times10^{38}$ erg s$^{-1}$)\\
(5)
\end{tabular}} &
\colhead{\begin{tabular}{c}
$L_{0.3-10\:keV}^{therm}$\\
($\times10^{38}$ erg s$^{-1}$)\\
(6)
\end{tabular}} &
\colhead{\begin{tabular}{c}
$\frac{L_{0.3-10\:keV}^{p.l.}}{L_{0.3-10\:keV}^{therm}}$\\
(7)
\end{tabular}} 
}
\startdata
Diff & \nodata & $34.61_{-5.40}^{+2.79}$ & $20.09_{-4.25}^{+9.15}$ & \nodata & $140.16_{-21.87}^{+11.34}$ & \nodata \\
Ring & $25.17_{13.54}^{+15.81}$ & $11.59_{-6.24}^{+7.32}$ & $0.93_{-0.18}^{+0.26}$ & $101.96_{-54.69}^{+64.00}$ & $46.95_{-25.12}^{+29.57}$ & 2.17 \\
Ring+ & $31.41_{-14.70}^{+10.91}$ & $13.63_{-6.39}^{+4.74}$ & $1.87_{-0.71}^{+0.87}$ & $127.23_{-59.54}^{+44.15}$ & $55.23_{-25.88}^{+19.20}$ & 2.30 \\
Inner Diff. & \nodata & $30.19_{-30.19}^{+23.76}$ & $18.87_{-8.54}^{+59.8}$ & \nodata & $122.31_{-122.31}^{+96.01}$ & \nodata \\
Outer Diff. & \nodata & $12.32_{-3.95}^{+2.34}$ & $6.37_{-1.20}^{+1.39}$ & \nodata & $49.91_{-16.00}^{+9.47}$ & \nodata \\
\enddata
\tablecomments{Col.(1)--Name of the extracted region. Col.(2)--The observed 0.3--10 keV flux of the power law component (if present). Col.(3)--The observed 0.3--10 keV flux of the thermal emission. Col.(4)--The emission measure defined as $EM=n^2V$.  Col.(5)--The absorption-corrected luminosity (0.3--10 keV) of the power law component. Col.(6)--The absorption-corrected luminosity of the thermal emission.  Col.(7)--The ratio between the intrinsic luminosities (0.3--10 keV) of the power law and the thermal emission.}
\end{deluxetable}

\clearpage


\begin{deluxetable}{cccccccc}
\rotate
\tabletypesize{\footnotesize}
\tablecaption{Hot-Gas Parameters.\label{mytable2}}
\tablewidth{0pt}
\tablehead{
\colhead{\begin{tabular}{c}
Region \\
(1)
\end{tabular}} &
\colhead{\begin{tabular}{c}
Area\\
(kpc$^2$)\\
(2)
\end{tabular}} &
\colhead{\begin{tabular}{c}
$n_e$\\
($\times10^{-2}$ cm$^{-3}$)\\
(3)
\end{tabular}} &
\colhead{\begin{tabular}{c}
$E_{th}$\\
($\times10^{54}$ erg)\\
(4)
\end{tabular}} &
\colhead{\begin{tabular}{c}
$\tau_c$\\
($\times10^{7}$ yr)\\
(5)
\end{tabular}} &
\colhead{\begin{tabular}{c}
$p$\\
($\times10^{-10}$ dyne cm$^{-2}$)\\
(6)
\end{tabular}} &
\colhead{\begin{tabular}{c}
$M_{ISM}$\\
($\times10^{5}$ M$_{\odot}$)\\
(7)
\end{tabular}} &
\colhead{\begin{tabular}{c}
$R_{SN}$\\
($10^{-3}$ yr$^{-1}$)\\
(8)
\end{tabular}}
}
\startdata
Diff. &  29.2 & 10.8 & 50.9 & 9.26 & 1.97 & 155.3 & 7.3\\
Ring &  1.1 & 11.9 & 2.3 & 8.64 & 2.34 & 6.5 & 0.3\\
Ring+ &  1.3 & 15.6 & 3.5 & 6.62 & 3.06 & 10.0 & 0.5\\
Inner Diff. &  5.8 & 23.5 & 18.9 & 3.95 & 3.69 & 67.1 & 2.7\\
Outer Diff. & 23.4 & 6.8 & 25.6 & 14.7 & 1.24 & 78.3 & 3.7\\
\enddata

\tablecomments{Col.(1)--Name of the extracted region. Col.(2)--Area of the extracted region (assuming a distance of $d\sim 18.6$ Mpc). Col.(3)--Thermal gas electron density derived from the emission measure ($EM=n^2V$). Col.(4)--Thermal energy contained in the gas.  Col.(5)--Cooling time of the hot gas, estimated following Tucker (1975). Col.(6)--Thermal pressure ($p=2nkT$) of the hot gas.  Col.(7)--Estimated mass of the hot ISM. Col.(8)--SN explosion rate needed to provide mechanical energy deposited to the ambient ISM (see Heckman et al. 1996, Fabbiano et al. 1997).}

\end{deluxetable}

\clearpage


\begin{deluxetable}{cccc}
\rotate
\tabletypesize{\footnotesize}
\tablecaption{Comparison of Relative Element Abundances.\label{mytable3}}
\tablewidth{0pt}
\tablehead{
\colhead{\begin{tabular}{c}
Region \\
(1)
\end{tabular}} &
\colhead{\begin{tabular}{c}
[Ne/Fe]\\
(2)
\end{tabular}} &
\colhead{\begin{tabular}{c}
[Mg/Fe]\\
(3)
\end{tabular}} &
\colhead{\begin{tabular}{c}
[Si/Fe]\\
(4)
\end{tabular}}
}
\startdata
All Diff & $0.05_{-0.33}^{+0.38}$ & $0.25_{-0.25}^{+0.28}$ & $0.40_{-0.32}^{+0.36}$\\
All Ring & $0.4_{-0.22}^{+0.38}$ & $0.34_{-0.24}^{+0.43}$ & $0.68_{-0.22}^{+0.40}$ \\
All Ring+ & $0.38_{-0.28}^{+1.12}$ & $0.35_{-0.27}^{+0.83}$ & $0.65_{-0.26}^{+0.75}$ \\
Inner Diff. & $0.23_{-0.40}^{+0.45}$ & $0.3_{-0.35}^{+0.47}$ & $0.56_{-0.39}^{+0.83}$ \\
Outer Diff. & $0.04_{-0.41}^{+0.62}$ & $0.35_{-0.29}^{+0.35}$ & $0.66_{-0.35}^{+0.44}$ \\
Hot halo emission S04\tablenotemark{a} & 0.44 & 0.49 & $<1.03$\\
Warm Galactic halo\tablenotemark{b} & \nodata & 0.25 & 0.5 \\
Type II SN A96\tablenotemark{c} & 0.19 & 0.37 & \nodata\\
Type II SN T95\tablenotemark{c} & 0.31 & 0.47 & 0.49\\
Type II SN W95A ($10^{-4}Z_{\odot}$)\tablenotemark{c} & 0.15 & 0.18 & 0.61\\
Type II SN W95B ($10^{-4}Z_{\odot}$)\tablenotemark{c} & 0.29 & 0.22 & 0.43 \\
Type II SN W95A ($Z_{\odot}$)\tablenotemark{c} & 0.24 & 0.25 & 0.49 \\
Type II SN W95B ($Z_{\odot}$)\tablenotemark{c} & 0.31 & 0.31 & 0.46\\
Type II SN N98S1\tablenotemark{d} & 0.41 & 0.61 & 0.63\\
Type I SN TNH93\tablenotemark{c} & -2.14 & -1.44 & -0.23\\
Type I SN WDD2\tablenotemark{d} & -2.8 & -1.7 & 0.01 \\
Type I SN W7\tablenotemark{d} & -2.18 & -1.42 & -0.25 \\
\enddata
\tablenotetext{(a)}{From Savage \& Sembach (1996).}
\tablenotetext{(b)}{From Strickland et al. (2004) Table 10.}
\tablenotetext{(c)}{IMF averaged stellar yields from Type I and II SNe compiled by Gibson et al. (1997). See Gibson et al. (1997) for model details and nomenclature.}
\tablenotetext{(d)}{Stellar yields from Type I and II SNe from Nakataki \& Sato (1998). See Nakataki \& Sato (1998) for model details and nomenclature.}

\end{deluxetable}

\clearpage

\begin{deluxetable}{llccccc}
\rotate
\tabletypesize{\small} 
\tablecaption{X-ray and FIR Luminosities of NGC 1365 and Other Starburst Galaxies\label{starburst}} 
\tablewidth{0pt}
\tablehead{\colhead{Galaxy} & \colhead{$t_{starburst}$ (Myr)\tablenotemark{a}}& \colhead{$\log N_H$ (cm$^{-2}$)\tablenotemark{b}} & \colhead{$L_{X,corr}$}(ergs s$^{-1}$)\tablenotemark{c}& \colhead{$L_{FIR}$(ergs s$^{-1}$)\tablenotemark{d}} & \colhead{$L_X$/$L_{FIR}$}&\colhead{References}
} 
\startdata
NGC 1365 & 3-6 & 20.5 & $5\times 10^{40}$ & $3.8\times 10^{44}$ & $1.3\times 10^{-4}$& 1,2,3\\
NGC 1365 Ring & 3-6 & 21.1 & $2.7\times 10^{40}$ & $1.4\times 10^{44}$ & $1.9\times 10^{-4}$& 1,4\\
M82 & 4-6 & 21.0 & $2\times 10^{40}$ & $9.3\times 10^{43}$& $2.2\times 10^{-4}$& 5,6,7\\
NGC 3351 & 4-5 & 20.7 & $3.3\times 10^{39}$ & $9.5\times10^{43}$& $3.5\times 10^{-5}$ & 8,9,3\\
Antennae & 4-13 & 21.2 & $2\times 10^{41}$ & $1.4\times 10^{44}$& $1.4\times 10^{-3}$ & 10,11,3\\
Arp 220 & 1-3 & 21.8 & $2.3\times 10^{41}$ & $6.8\times 10^{45}$& $3.4\times 10^{-5}$ & 12,13,14\\
NGC 253 & 20-30 & 22.0 & $6\times 10^{39}$ & $7\times 10^{43}$& $9\times 10^{-5}$ & 15,16,17\\
Galactic Center & 2-7 & 22.8 & $10^{38}:$ & $3.8\times 10^{42}$& $3\times 10^{-5}$& 18,19\\
\enddata

\tablenotetext{a}{The age of recent starburst event is taken as the measured range of young star clusters in the starburst galaxy (see Gallagher \& Smith 2005 review on difficulties in measuring the ages of starbursts). In the Galactic center, Figer (2008) shows 2--3 Myr for the Arches and the Quintuplet clusters, 3--7 Myr for the Central cluster.}

\tablenotetext{b}{$N_H$ values are taken from X-ray spectral fitting reported in literature if available, otherwise we use $A_V\approx 4.5\times 10^{-22} N_H$ (mag) (Ryter 1996) to estimate $N_H$. For Arp 220 and NGC 253, we note that both their nuclear starburst regions are heavily obscured with $N_H\ga 10^{23}$ cm$^{-2}$; the values shown here are typical of the more extended regions. Therefore, the absorption corrected luminosities most likely lower than the intrinsic soft X-ray luminosities.}

\tablenotetext{c}{The absorption corrected soft X-ray luminosity is taken from {\em ROSAT} results (0.1--2.5 keV) reported in the literature and this work, except for NGC 3351 and the Galactic center.  The $L_X$ for NGC 3351 is converted from {\em Chandra} measurement using the best fit model from Swartz et al. (2006).  Because of the heavy obscuration, only 2--8 keV measurement is available for the Galactic center and the absorption corrected $L_{X,corr}$ is highly uncertain. For the galaxies with $\log N_H\sim 21$, the uncertainty in the absorption columns results in a $\pm$0.08 dex uncertainty in the absorption corrected $L_{X,corr}$ from the XSPEC spectral fit.  For Arp 220 and NGC 253, correcting for a $\log N_H\sim 23$ column (the most obscured nuclear region) can increase $L_{X,corr}$ by a factor of 100.  For comparison, we list here X-ray luminosities (0.2--4.0 keV) from Fabbiano et al. (1992) with the {\em Einstein Observatory} measurements--NGC 1365: $\log Lx=40.84$, M82: $\log Lx= 40.56$, NGC 3351: $\log Lx<39.98$, NGC 4038/9: (Antennae) $\log Lx=40.96$, NGC 253: $\log Lx=39.87$, Arp 220: $\log Lx= 41.68$.}

\tablenotetext{d}{The FIR luminosities (except that of the NGC 1365 ring) are derived using IRAS measurements ($\sim$10\% uncertainty; Ranalli et al. 2003). Helou \& Soifer (1985) gives $FIR=1.26\times 10^{-14}(2.58f_{60\mu m}+f_{100\mu m})$ W m$^{-2}$, which is commonly used in the references, where $f_{\nu}$ are the flux densities in Jy.  FIR emission from the NGC 1365 ring cannot be accurately measured due to the low spatial resolution at 70$\mu$m and 160$\mu$m (saturated) {\em Spitzer} images.  We estimate $L_{FIR}$ of the ring is $\sim$35\% of the IRAS $L_{FIR}$ for NGC 1365.}

\tablerefs{1. Galliano et al. (2005, 2008); 2. Stevens et al.(1999); 3. Sanders et al. (2003); 4. This work; 5. Smith et al. (2006); 6. Strickland, Ponman \& Stevens (1997) 7. Condon et al.(1998); 8. Colina et al. (1997); 9. Swartz et al.(2006); 10. Mengel et al. (2001); 11. Read et al. (1995), see also Fabbiano et al. (1997); 12. Wilson et al.(2006); 13. Heckman et al. (1996), see also McDowell et al. (2003); 14. Soifer et al. (1984); 15. Engelbracht et al. (1998); 16. Weaver et al. (2002); 17. Strickland et al. (2004); 18. Figer (2008) and references therein; 19. Morris \& Serabyn (1996); see also Wang et al. (2006).}

\end{deluxetable}

\end{document}